\documentclass{elsarticle}
\usepackage{epstopdf}
\usepackage{hyperref}
\usepackage[caption=false,font=footnotesize]{subfig}
\usepackage{tikz}
\usepackage{booktabs}
\usepackage{amssymb}
\usepackage{amsmath}

\usepackage{wrapfig}
\usetikzlibrary{arrows,patterns,automata,backgrounds,decorations,fit,petri,positioning,petri,shapes,calc}
\tikzset{place/.append style={circle,draw=black,thick,inner sep=0pt,minimum size=3mm,label position=below}}
\tikzset{transition/.append style={rectangle,draw=black,thick,inner sep=1pt,minimum size=7mm}}
\tikzset{every edge/.append style={-{>[sep=0pt]}, thin}}
\tikzset{pre/.append style={<-,shorten <=0pt,shorten >=0pt}}
\tikzset{post/.append style={->,shorten >=0pt,shorten <=0pt}}
\newtheorem{definition}{Definition}

\newcommand{\fleche}{\longrightarrow}
\newcommand{\flsup}[1]{\stackrel{#1}{\fleche}}
\newcommand{\step}[1]{\flsup{#1}}           
\newcommand{\Lan}                 {\mathfrak{L}}

\newcommand{\pre}[1]{\bullet #1}
\newcommand{\APN}{\mathit{APN}}
\newcommand{\LN}{\mathit{LN}}
\newcommand{\MF}{\mathit{MF}}

\newlength{\hatchspread}
\newlength{\hatchthickness}
\newlength{\hatchshift}
\newcommand{\hatchcolor}{}
\tikzset{hatchspread/.code={\setlength{\hatchspread}{#1}},
	hatchthickness/.code={\setlength{\hatchthickness}{#1}},
	hatchshift/.code={\setlength{\hatchshift}{#1}},
	hatchcolor/.code={\renewcommand{\hatchcolor}{#1}}}
\tikzset{hatchspread=3pt,
	hatchthickness=0.4pt,
	hatchshift=0pt,
	hatchcolor=black}
\pgfdeclarepatternformonly[\hatchspread,\hatchthickness,\hatchshift,\hatchcolor]
{custom north west lines}
{\pgfqpoint{\dimexpr-2\hatchthickness}{\dimexpr-2\hatchthickness}}
{\pgfqpoint{\dimexpr\hatchspread+2\hatchthickness}{\dimexpr\hatchspread+2\hatchthickness}}
{\pgfqpoint{\dimexpr\hatchspread}{\dimexpr\hatchspread}}
{
	\pgfsetlinewidth{\hatchthickness}
	\pgfpathmoveto{\pgfqpoint{0pt}{\dimexpr\hatchspread+\hatchshift}}
	\pgfpathlineto{\pgfqpoint{\dimexpr\hatchspread+0.15pt+\hatchshift}{-0.15pt}}
	\ifdim \hatchshift > 0pt
	\pgfpathmoveto{\pgfqpoint{0pt}{\hatchshift}}
	\pgfpathlineto{\pgfqpoint{\dimexpr0.15pt+\hatchshift}{-0.15pt}}
	\fi
	\pgfsetstrokecolor{\hatchcolor}
	\pgfusepath{stroke}
}

\journal{Journal of Innovation in Digital Ecosystems}

\begin{document}

\begin{frontmatter}

\title{Mining Local Process Models}

\author[a,b]{Niek Tax}
\author[a]{Natalia Sidorova}
\author[b]{Reinder Haakma}
\author[a]{Wil M. P. van der Aalst}

\address[a]{Eindhoven University of Technology, P.O. Box 513, Eindhoven, The Netherlands}
\address[b]{Philips Research, Prof. Holstlaan 4, 5665 AA Eindhoven, The Netherlands}

\begin{abstract}
In this paper we describe a method to discover frequent behavioral patterns in event logs. We express these patterns as \emph{local process models}. Local process model mining can be positioned in-between process discovery and episode / sequential pattern mining. The technique presented in this paper is able to learn behavioral patterns involving sequential composition, concurrency, choice and loop, like in process mining. However, we do not look at start-to-end models, which distinguishes our approach from process discovery and creates a link to episode / sequential pattern mining. We propose an incremental procedure for building local process models capturing frequent patterns based on so-called process trees. We propose five quality dimensions and corresponding metrics for local process models, given an event log. We show monotonicity properties for some quality dimensions, enabling a speedup of local process model discovery through pruning. We demonstrate through a real life case study that mining local patterns allows us to get insights in processes where regular start-to-end process discovery techniques are only able to learn unstructured, flower-like, models.
\end{abstract}


\end{frontmatter}

\section{Introduction}
Process mining aims to extract novel insight from event data \cite{Aalst2016}. Process discovery, the task of discovering a process model that is representative for the set of event sequences in terms of start-to-end behavior, i.e. from the start of a case till its termination, plays a prominent role in process mining. Many process discovery algorithms have been proposed and applied to a variety of real life cases (see \cite{Aalst2016} for an overview). A different perspective on mining patterns in event sequences can be found in the data mining field, where the episode mining \cite{Mannila1997} and sequential pattern mining \cite{Agrawal1995} subfields focus on finding frequent patterns that are local, not necessarily describing the whole event sequences from start to end. Episode mining and sequential pattern mining have e.g. been used to analyze telecommunication networks \cite{Mannila1997}, web navigational logs \cite{Mannila1997,Casas-Garriga2003}, and retail sales transactions \cite{Atallah2004}.

Sequential pattern mining and episode mining are limited to the discovery of \emph{sequential orderings} or \emph{partially ordered sets} of events, while process discovery methods aim to discover a larger set of event relations, including sequential orderings, (exclusive) choice relations, concurrency, and loops, represented in process models such as Petri nets \cite{Reisig2012}, BPMN \cite{OMG2011}, and process trees \cite{Buijs2014}. Process models that can be discovered with process discovery methods distinguish themselves from more traditional sequence mining methods like Hidden Markov Models \cite{Rabiner1989} and Recurrent Neural Networks \cite{Goller1996,Hochreiter1997} in that process models can be visually represented and their visual representation can be used for communication between process stakeholders. However, process discovery is normally limited to the discovery of a model capturing the behavior of process instances as a whole, and not local patterns within instances. Our goal is to develop methods allowing to mine \emph{local} process models positioned in-between simple patterns (e.g. subsequences) and start-to-end models. Local process models focus on a subset of the process activities and describe some behavioral pattern that occurs frequently within event sequences. Such local process models cannot be discovered by using standard techniques.

\begin{figure*}[t]
	\centering
	\hspace{-0.3cm}
	\scalebox{0.75}{
		\subfloat[]{
			\begin{tabular}{l}
				\toprule
				Event sequences \\
				\midrule
				$\langle$A,\textbf{A,C,B},A,\textbf{A,C,B},B,C$\rangle$\\
				$\langle$C,\textbf{A,C,B},A,A,\textbf{A,B,C},B$\rangle$\\
				$\langle$A,\underline{\textbf{A},\textbf{B},D,\textbf{C}},D,\textbf{A,B,C},B$\rangle$\\
				$\langle$C,\textbf{A,C,B},B,B,\underline{\textbf{A},D,\textbf{B,C}}$\rangle$\\
				$\langle$B,\textbf{A,B,C},C$\rangle$\\
				$\langle$D,\textbf{A,C,B},C,A,\underline{\textbf{A,C},A,\textbf{B}}$\rangle$\\
				$\langle$D,\textbf{A,B,C},D,C,A,C,\textbf{A,B,C}$\rangle$\\
				\bottomrule
			\end{tabular}
			\label{sfig:traces_example}
		}
		\hspace{-0.35cm}
		\subfloat[]{
			\scalebox{0.89}{
				\raisebox{-.5\height}{
					\begin{tikzpicture}
					[node distance=1.2cm,
					on grid,>=stealth',
					bend angle=20,
					auto,
					every place/.style= {minimum size=5mm},
					every transition/.style = {minimum size = 5mm},
					transitionH/.style={rectangle, thick, fill=black, minimum width=3mm, inner ysep=9pt }
					]
					\node [place, tokens = 1] (p){};
					\node [transition] (2) [align=center, right = of p] {\textbf{A} 13/21}
					edge [pre] node[auto] {} (p);
					\node [place] (p3) [above right = of 2] {}
					edge[pre] node[auto] {} (2);
					\node [transition] (t1) [right = of p3]{\textbf{B} 13/20}
					edge[pre] node[auto] {} (p3);
					\node [place] (p4) [below right = of 2] {}
					edge[pre] node[auto] {} (2);
					\node [transition] (t2) [right = of p4] {\textbf{C} 13/19}
					edge[pre] node[auto] {} (p4);
					\node [place] (p5) [right = of t1] {}
					edge[pre] node[auto] {} (t1);
					\node [place] (p6) [right = of t2] {}
					edge[pre] node[auto] {} (t2);
					\node [transitionH] (t3) [below right = of p5] {}
					edge[pre] node[auto] {} (p5)
					edge[pre] node[auto] {} (p6);
					\node [place] (p7) [right=of t3] {}
					edge[pre] node[auto] {}(t3);
					\end{tikzpicture}}}
			\label{sfig:local_model_example}
		}
		\hspace{-0.27cm}
		\subfloat[]{
			\raisebox{-.5\height}{
				\scalebox{0.89}{
					\begin{tikzpicture}
					[node distance=1.1cm,
					on grid,>=stealth',
					bend angle=20,
					auto,
					every place/.style= {minimum size=5mm},
					every transition/.style = {minimum size = 5mm},
					transitionH/.style={rectangle, thick, fill=black, minimum width=3mm, inner ysep=9pt }
					]
					\node [place, tokens = 1] (p){};
					\node [transitionH] (2) [right = of p] {}
					edge [pre] node[auto] {} (p);
					\node [place] (p3) [right = of 2] {}
					edge[pre] node[auto] {} (2);
					\node [transition] () [label=center:B, above right = of p] {}
					edge [pre, bend left] node[auto] {} (p)
					edge [post, bend right] node[auto] {} (p);
					\node [transition] () [label=center:A, above = of p] {}
					edge [pre, bend left] node[auto] {} (p)
					edge [post, bend right] node[auto] {} (p);
					\node [transition] () [label=center:C, below = of p] {}
					edge [pre, bend left] node[auto] {} (p)
					edge [post, bend right] node[auto] {} (p);
					\node [transition] () [label=center:D, below right = of p] {}
					edge [pre, bend left] node[auto] {} (p)
					edge [post, bend right] node[auto] {} (p);
					\end{tikzpicture}}}
			\label{sfig:flower_example}
		}
		\hspace{-0.27cm}
		\subfloat[]{
			\scalebox{0.9}{
				\begin{tabular}{l}
					\toprule
					Sequential\\ patterns \\
					\midrule
					$\langle A,B,A\rangle$\\
					$\langle A,B,C\rangle$\\
					$\langle A,C,A\rangle$\\
					$\langle A,C,B\rangle$\\
					$\langle B,A,B\rangle$\\
					$\langle B,A,C\rangle$\\
					$\langle C,A,C\rangle$\\
					\bottomrule
				\end{tabular}
				\label{sfig:sequential_patterns}
			}
		}
	}
	
	\caption{\emph{(a)} A log $L$ of sales officer event sequences with highlighted instances of the frequent pattern. \emph{(b)} The local process model showing frequent behavior in $L$. \emph{(c)} The Petri net discovered on $L$ with the Inductive Miner algorithm \cite{Leemans2013}. \emph{(d)} The sequential patterns discovered on $L$ with the PrefixSpan algorithm \cite{Pei2001}.}
	\label{fig:unstructured_log_discovery}
	\vspace{-0.1cm}
\end{figure*}
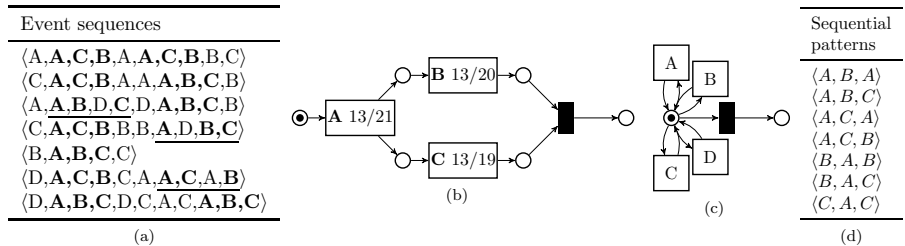

Imagine a sales department where multiple sales officers perform four types of activities: (A) register a call for bids, (B) investigate a call for bids from the business perspective, (C) investigate a call for bids from the legal perspective, and (D) decide on participation in the call for bid. The event sequences (Figure \ref{sfig:traces_example}) contain the activities performed by one sales officer throughout the day. The sales officer works on different calls for bids and not necessarily performs all activities for a particular call himself. Applying discovery algorithms, like the Inductive Miner \cite{Leemans2013}, yields models allowing for any sequence of events (Figure \ref{sfig:flower_example}). Such "flower-like" models do not give any insight in typical behavioral patterns. When we apply any sequential pattern mining algorithm using a threshold of six occurrences, we obtain the seven length-three sequential patterns depicted in Figure \ref{sfig:sequential_patterns} (results obtained using the SPMF \cite{Fournier2014} implementation of the PrefixSpan algorithm \cite{Pei2001}). However, the data contains a frequent non-sequential pattern where a sales officer first performs \emph{A}, followed by \emph{B} and a \emph{C} in arbitrary order (Figure \ref{sfig:local_model_example}). This pattern cannot be found with existing process discovery and sequential pattern mining techniques. The two numbers shown in the transitions (i.e., rectangles) represent (1) the number of events of this type in the event log that fit this local process model and (2) the total number of events of this type in the event log. For example, 13 out of 19 events of type \emph{C} in the event log fit transition \emph{C}, which are indicated in bold in the log in Figure \ref{sfig:traces_example}. Underlined sequences indicate non-continuous instances, i.e. instances with non-fitting events in-between the events forming the instance of the local process model.

In this paper we describe a method to extract frequently occurring \emph{local process models}, allowing for choice, concurrency, loops, and sequence relations. We leverage process trees \cite{Buijs2012} to search for local process models, and describe a way to recursively explore candidate process trees up to a certain model size. For convenience, we often use the Petri net representations for process trees. In fact, results can also be visualized as BPMN \cite{OMG2011}, EPC \cite{Keller1992}, UML activity diagram \cite{ISO2012}, or UML statechart diagram \cite{ISO2012}. We define five quality dimensions that express the degree of representativeness of a local process model with regard to an event log: \emph{support}, \emph{confidence}, \emph{language fit}, \emph{coverage}, and \emph{determinism}. Based on quality metrics, we describe monotonicity properties over some quality dimensions and show how they can be used to make the recursive search over process trees more efficient.\looseness=-1

The paper is organized as follows. Section \ref{sec:preliminaries} introduces the basic concepts used in this paper. Section \ref{sec:related} describes related work in the fields of process discovery and sequential pattern mining. Section \ref{sec:approach} describes our local process model mining approach. Section \ref{sec:quality} introduces quality dimensions and metrics for local process models and discusses monotonicity properties. Section \ref{sec:alignment_based_evaluation} describes a local process model evaluation approach based on alignments. Section \ref{sec:case_studies} shows the relevance of the proposed technique using two real life data sets and compares the results with the results obtained with several related techniques. Section \ref{sec:conclusion} concludes the paper.

\section{Preliminaries}
\label{sec:preliminaries}
In this section we introduce process modeling notations, including Petri nets, process trees, which are used in later sections of this paper.

$X^*$ denotes the set of all sequences over a set $X$ and $\sigma=\langle a_1,a_2,\dots,a_n\rangle$ a sequence of length $n$; $\langle\rangle$ is the empty sequence and $\sigma_1 \cdot \sigma_2$ is the concatenation of sequences $\sigma_1,\sigma_2$. $\sigma\upharpoonright Q$ is the projection of $\sigma$ on $Q$, e.g. $\langle a,b,c,a,b,c\rangle\upharpoonright_{\{a,c\}}=\langle a,c,a,c \rangle$. $\#_a(\sigma)$ denotes the number of occurrences of element $a$ in sequence $\sigma$, e.g. $\#_a(\langle a,b,c,a\rangle)=2$.

\begin{definition}[Applying Functions to Sequences]
	\label{def:funtoseq}
	A partial function $f \in X \nrightarrow Y$ can be lifted to sequences over $X$ using the following recursive definition: (1) $f(\langle\rangle)=\langle\rangle$;  (2) for any $\sigma\in X^*$ and $x\in X$:
	\begin{center}
		$f(\sigma \cdot \langle x\rangle) =
		\left\{
		\begin{array}{ll}
		f(\sigma)  & \mbox{if } x \notin dom(f), \\
		f(\sigma) \cdot \langle f(x)\rangle & \mbox{if } x \in dom(f).
		\end{array}
		\right.$
	\end{center}
\end{definition}

We assume the set of all \emph{process activities} $\Sigma_L$ to be given. An \emph{event} $e$ is the occurrence of an activity $e\in \Sigma_L$. We call a sequence of events $t\in {\Sigma_L}^*$ a \emph{trace}. An \emph{event log} $L \in \mathbb{N}^{{\Sigma_L}^*}$ is a finite multiset of traces. For example, the event log $L=[\langle a,b,c\rangle^2,\langle b,a,c\rangle^3]$ consists of 2 occurrences of trace $\langle a,b,c\rangle$ and three occurrences of trace $\langle b,a,c\rangle$. We lift the sequence projection to the multisets of sequences in the standard way. For example, for the log $L$ given above $L\upharpoonright_{\{a,c\}}=[\langle a,c\rangle^5]$. We lift the number of occurrences in a sequence to multisets of sequences in the standard way, for example, $\#_a(L)=5$.\looseness=-1

Petri nets are directed bipartite graphs consisting of transitions and places, connected by arcs. Transitions represent activities, while places represent the enabling conditions of transitions. Labels are assigned to transitions to indicate the type of activity that they model. A special label $\tau$ is used to represent invisible transitions, which are only used for routing purposes and not recorded in the execution log.
\begin{definition}[Labeled Petri net]
	\label{def:lpn}
	A \emph{labeled Petri net} $N=\langle P,T,F,\Sigma_M,\ell\rangle$ is a tuple where $P$ is a finite set of places, $T$ is a finite set of transitions such that $P \cap T = \emptyset$,  $F \subseteq (P \times T) \cup (T \times P)$ is a set of directed arcs, called the flow relation, $\Sigma_M$ is a finite set of labels representing activities, with $\tau \notin \Sigma_M$ being a label representing  invisible events, and $\ell:T\rightarrow \Sigma_M\cup \{\tau\}$ is a labeling function that assigns a label to each transition.
	\end{definition}
For a node $n \in (P \cup T)$ we use $\bullet n$ and $n \bullet$ to denote the set of input and output nodes of $n$, defined as $\bullet n =\{n'|(n',n)\in F\}$ and $n \bullet =\{n|(n,n')\in F\}$ . 

A state of a Petri net is defined by its \emph{marking} $M \in \mathbb{N}^{P}$ being a multiset of places. A marking is graphically denoted by putting $M(p)$ tokens on each place $p\in P$. A pair $(N,M)$ is called a marked Petri net. State changes occur through transition firings. A transition $t$ is enabled (can fire) in a given marking $M$ if each input place $p\in \pre{t}$ contains at least one token. Once a transition fires, one token is removed from each input place of $t$  and one token is added to each output place of $t$, leading to a new marking $M'$ defined as $M'=M-\bullet t+t\bullet$.
A firing of a transition $t$ leading from marking $M$ to marking $M'$ is denoted as $M \step{l(t)} M'$. $M_1 \step{l(\sigma)} M_2$ indicates that $M_2$ can be reached from $M_1$ through a firing sequence $\sigma'\in {\Sigma_M}^*$.\looseness=-1

Often it is useful to consider a Petri net in combination with an initial marking and a set of possible final markings. This allows us to define the language accepted by the Petri net and to check whether some  behavior is part of the behavior of the Petri net (can be replayed on it).

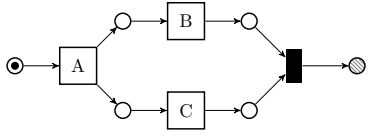
\begin{figure}[t]
\centering
\scalebox{0.7}{
\begin{tikzpicture}
[node distance=1.2cm,
on grid,>=stealth',
bend angle=20,
auto,
every place/.style= {minimum size=5mm},
every transition/.style = {minimum size = 5mm},
transitionH/.style={rectangle, thick, fill=black, minimum width=3mm, inner ysep=9pt }
]
\node [place, tokens = 1] (p){};
\node [transition] (2) [align=center, right = of p] {A}
edge [pre] node[auto] {} (p);
\node [place] (p3) [above right = of 2] {}
edge[pre] node[auto] {} (2);
\node [transition] (t1) [right = of p3]{B}
edge[pre] node[auto] {} (p3);
\node [place] (p4) [below right = of 2] {}
edge[pre] node[auto] {} (2);
\node [transition] (t2) [right = of p4] {C}
edge[pre] node[auto] {} (p4);
\node [place] (p5) [right = of t1] {}
edge[pre] node[auto] {} (t1);
\node [place] (p6) [right = of t2] {}
edge[pre] node[auto] {} (t2);
\node [transitionH] (t3) [below right = of p5] {}
edge[pre] node[auto] {} (p5)
edge[pre] node[auto] {} (p6);
\node [place,pattern=custom north west lines,hatchspread=1.5pt,hatchthickness=0.25pt,hatchcolor=gray] (p7) [right=of t3] {}
edge[pre] node[auto] {}(t3);
\end{tikzpicture}}

\caption{Example of an accepting Petri net.}
\label{example_apn}
\end{figure}

\begin{definition}[Accepting Petri net]
An \emph{accepting Petri net} is a triple $\APN = (N,M_0,\MF)$, where $N$ is a labeled Petri net, $M_0\in\mathbb{N}^p$ is its initial marking, and $\MF\subseteq\mathbb{N}^p$ is its set of possible final markings, such that $\forall_{M_1,M_2 \in \MF} \;M_1\nsubseteq M_2$. A sequence $\sigma\in T^*$ is called a \emph{trace} of an accepting Petri net $\APN$ if $M_0\step{\sigma} M_f$ for some final marking $M_f\in\MF$. The \emph{language} $\Lan(APN)$ of $\APN$ is the set of all its traces.
\end{definition}

Figure \ref{example_apn} shows an example of an accepting Petri net. Circles represent places and rectangles represent transitions. Invisible transitions (labeled $\tau$) are depicted as black rectangles. Places that belong to the initial marking contain a token and places belonging to a final marking contain a bottom right label $f_i$ with $i$ a final marking identifier, or are simply marked as $\begin{tikzpicture}
[node distance=1.4cm,
on grid,>=stealth',
bend angle=20,
auto,
every place/.style= {minimum size=0.1mm},
]
\node [place,pattern=custom north west lines,hatchspread=1.5pt,hatchthickness=0.25pt,hatchcolor=gray] {};
\end{tikzpicture}$ in case of a single final marking. The language of this accepting Petri net is $\{\langle A,B,C\rangle,\langle A,C,B\rangle\}$.

A different process representation is a process tree~\cite{Buijs2012}. Process trees can only model sound (deadlock-free and livelock-free) processes. The recursive definition of process trees make them a convenient representation to iteratively expand  process models into larger  process models.

\begin{definition}[Process tree]
Let $A\in\Sigma_M$ be a finite set of activities with $\tau \notin \Sigma_M$. $\bigoplus=\{\rightarrow,\times,\wedge,\circlearrowright\}$ is the set of \emph{process tree} operators.
\begin{itemize}
	\item if $a\in \Sigma_M \cup \{\tau\}$ then $Q=a$ is a process tree.
	\item if $Q_1,Q_2$ are process trees, and $\oplus\in\bigoplus$, then $Q=\oplus(Q_1,Q_2)$ is a process tree.\looseness=-1
\end{itemize}
\end{definition}
A \emph{process tree} is a tree structure consisting of operator and activity nodes, such that each leaf node is an activity node and each non-leaf node is an operator node.\looseness=-1

The \emph{loop} operator ($\circlearrowright$) has two child nodes, with the first child the "do" part and the second child the "redo" child. Process tree $p_1=\mathit{\circlearrowright(a,b)}$ accepts language $\Lan(p_1)=\{\langle a \rangle, \langle a,b,a, \rangle, \langle a,b,a,b,a \rangle,\dots\}$.

The \emph{sequence} operator ($\rightarrow$) has two children, such that the first child is executed prior to execution of the second child. The language of process tree $p_2=\mathit{\rightarrow(a,b)}$ is $\Lan(p_2)=\{\langle a,b \rangle\}$.

The \emph{exclusive choice} operator ($\times$) has two children, indicating that either the first or the second child will be executed, but not both. The language of process tree $p_3=\times(a,b)$ is $\Lan(p_3)=\{\langle a \rangle,\langle b \rangle\}$.

The \emph{concurrency} operator ($\wedge$) has two children, indicating that the both children will be executed in parallel. Process tree $p_4=\mathit{\wedge(\rightarrow(a,b),\rightarrow(c,d))}$ accepts language $\Lan(p_4)=\{\langle a,b,c,d \rangle,\langle a,c,b,d \rangle,\langle a,c,d,b \rangle,\langle c,a,b,d \rangle, \langle c,a,d,b \rangle,\\\langle c,d,a,b \rangle\}$.

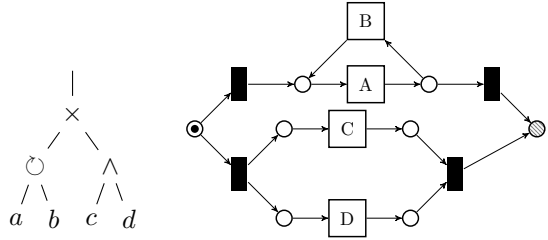
\begin{figure}[t]
\centering
\subfloat{
\begin{tikzpicture}[level distance=7mm]
\tikzstyle{level 1}=[sibling distance=20mm]
\tikzstyle{level 2}=[sibling distance=10mm]
\tikzstyle{level 3}=[sibling distance=5mm]
\node(9){}{
	child{node(5){$\times$} 
		child{node(B0) {$\circlearrowright$}
			child{node(Q){$a$}}
			child{node(W){$b$}}
		}
		child{node(R) {$\wedge$}
			child{node(Q){$c$}}
			child{node(W){$d$}}
		}
	}
};

\end{tikzpicture}}
\quad
\subfloat{
\scalebox{0.7}{
	\begin{tikzpicture}
	[node distance=1.2cm,
	on grid,>=stealth',
	bend angle=20,
	auto,
	every place/.style= {minimum size=5mm},
	every transition/.style = {minimum size = 5mm},
	transitionH/.style={rectangle, thick, fill=black, minimum width=3mm, inner ysep=9pt }
	]
	\node [place, tokens = 1] (p){};
	\node [transitionH] (s1) [align=center, above right = of p] {}
	edge [pre] node[auto] {} (p);
	\node [place] (p2) [align=center, right=of s1]{}
	edge [pre] node[auto] {} (s1);
	\node [transition] (t1) [align=center, right = of p2]{A}
	edge [pre] node[auto] {} (p2);
	\node [place] (p3) [right = of t1] {}
	edge [pre] node[auto] {} (t1);
	\node [transition] (t2) [above = of t1] {B}
	edge [pre] node[auto] {} (p3)
	edge [post] node[auto] {} (p2);
	\node [transitionH] (s2) [right = of p3]{}
	edge [pre] node[auto] {} (p3);
	\node [transitionH] (s3) [align=center, below right = of p] {}
	edge [pre] node[auto] {} (p);
	\node [place] (p4) [above right = of s3] {}
	edge [pre] node[auto] {} (s3);
	\node [transition] (t3) [align=center, right = of p4] {C}
	edge [pre] node [auto] {} (p4);
	\node [place] (p6) [right = of t3] {}
	edge [pre] node[auto] {} (t3);
	\node [place] (p5) [below right = of s3] {}
	edge [pre] node[auto] {} (s3);
	\node [transition] (t4) [align=center, right = of p5] {D}
	edge [pre] node [auto] {} (p5);
	\node [place] (p7) [right = of t4] {}
	edge [pre] node[auto] {} (t4);
	\node [transitionH] (s4) [below right = of p6]{}
	edge [pre] node[auto] {} (p6)
	edge [pre] node[auto] {} (p7);
	\node [place,pattern=custom north west lines,hatchspread=1.5pt,hatchthickness=0.25pt,hatchcolor=gray] (p7) [below right=of s2] {}
	edge[pre] node[auto] {} (s2)
	edge[pre] node[auto] {} (s4);
	\end{tikzpicture}
}
}
\caption{Graphical representation of process tree $\times(\circlearrowright(a,b),\wedge(c,d))$.}
\label{fig:process_tree}
\end{figure}

Let $p_5=\times(\circlearrowright(a,b),\wedge(c,d))$. Figure \ref{fig:process_tree} shows the graphical representation of $p_5$. Because the do and redo children of a loop operator can be executed an arbitrary number of times, process trees containing a loop operator have a language of infinite size. 

We define \emph{$n$-restricted language}, consisting of all language traces of at most length $n\in\mathbb{N}$, as $\Lan_n=\{t\in\Lan\;| \; |t| \le n\}$. The $n$-restricted language is guaranteed to be of finite size, independent of the operators in the process tree. The \mbox{$5$-restricted} language of the process tree $p_5$ is $\Lan_5(p_5)=\{\langle a\rangle,\langle a,b,a\rangle,\langle a,b,a,b,a\rangle,\\\langle c,d\rangle,\langle d,c\rangle\}$.
\\



\section{Related Work}
\label{sec:related}
ProM's Episode Miner \cite{Leemans2014} is a method that can be considered to be in-between episode mining and process mining, as it discovers a collection of patterns from an event log where each pattern consists of partial order constructs. However, contrary to the technique that we describe in this paper, ProM's Episode Miner does not support loop and exclusive choice constructs and is not easily extensible to include new types of constructs.

Lu et al. propose a method called Post Sequential Patterns Mining (PSPM) \cite{Lu2009} that takes a input a set of sequential patterns and post-processes them into a single graph consisting of sequential and exclusive choice constructs, which they call a Sequential Pattern Graph (SGP) \cite{Lu2004}. A later extension by Lu et al. adds the capability to mine concurrency relations \cite{Lu2011}. An SGP can be discovered from an event log by first applying any existing sequential pattern mining algorithm followed by PSPM on the discovered set of sequential patterns. The reliance of PSPM methods on the output of sequential pattern mining techniques can also be considered to be a drawback of the approach. When two activities $A$ and $B$ are in parallel then both the orderings $\langle A,B\rangle$ and $\langle B,A\rangle$ will be present in the complete log. However, if one of the two orderings is more frequent than the other due to chance, one of the two orderings might not reach the support set in the sequential pattern mining, making it impossible for PSPM methods to discover the concurrency of $A$ and $B$. A more fundamental distinction between PSPM and Local Process Model (LPM) discovery is that PSPM merges all relations into one single pattern while LPM discovery aims at discovering a collection of patterns. Merging separate patterns into a single graph could result in one single overgeneralizing graph. For example in the case of log $L=[\langle b,a,c\rangle^{100}, \langle d,a,e\rangle^{100}]$, sequential pattern mining techniques will find two sequential patterns: $b,a,c$ and $d,a,e$. Merging them into a single graph where $a$ is followed by either $c$ or $e$ and is preceded by either $b$ or $d$ results in the loss the long term dependency where $b$ already determines the occurrence of a $c$ after the $a$.\looseness=-1

Jung et al. \cite{Jung2008} describe a method to mine frequent patterns from a collection of process models by transforming each business process to a vector format and then applying agglomerative clustering. Diamantini et al. \cite{Diamantini2012,Diamantini2013} take a similar approach, but apply graph clustering techniques instead of a traditional clustering approach. These techniques differ from LPM discovery as they take as input a set of process models instead of an event log. However, in many situations there are no existing process models available and, as shown in the introduction, it is not always possible to mine structured process models from an event log using process discovery techniques.

In later work, Diamantini et al. \cite{Diamantini2016} describe a method to mine frequent patterns in process model notation through a two step approach. First each trace from the event log is transformed into a so called instance graph \cite{Dongen2004}, which is graph representation of a trace that shows which steps in the trace are performed sequentially and which steps are performed in parallel (i.e. overlapping in time). In the second step they apply a graph clustering technique to obtain frequent subgraphs from this set of instance graphs. However, since instance graphs are limited to sequential and parallel constructs, other process constructs, such as choices and loops, cannot be discovered with the approach described in Diamantini et al. \cite{Diamantini2016}, while they can be discovered with LPM discovery.

The techniques developed in the area of trace clustering \cite{Song2008,Bose2009,Folino2011,Hompes2014} are related to LPM discovery in the sense that both aim to enable extraction of process insight from event logs where the process is too unstructured for existing process discovery techniques to find a structured process model. Trace clustering techniques aim to solve this by clustering similar traces together to prevent mixing different usage scenarios into one single unstructured process model. Trace clustering techniques work well when the original event log does not originate from one single process, but in fact originates from multiple processes. However, not in all types of complex and flexible event data there is a cluster tendency in the data. An example for such non-clusterable event data can be found in the log shown in Figure \ref{sfig:traces_example}, where no clustering over the traces would enable the discovery of the frequent pattern shown in Figure \ref{sfig:local_model_example}. The traces in the log have large parts of randomness \emph{within} the traces, while trace clustering helps for cases where there is a large degree of variety \emph{between} traces.

Declarative process models, such as Declare \cite{pesic2007}, define the allowed behavior through constraints that must be respected while carrying out the process. This contrasts procedural process models, which are dominant in the process discovery field and specify all possible orderings of events explicitly. Two examples of process discovery approaches that generate declarative process models are the DPIL Miner \cite{Schonig2015} and the Declare Miner \cite{Maggi2011}. Both approaches specify a set of \emph{rule templates} that consists of two activity variables and their relation. An example of such a template is $\mathit{sequence}(a,b)$, indicating that some activity $a$ is followed by $b$. Concrete rules are extracted from the event log based on this template-based search. However, since the rule templates are limited to relations between two activities, more complex relations between three or more activities cannot be discovered. Imagine that for some event log a declarative process discovery method finds two relations: $\mathit{sequence}(a,b)$ and $\mathit{sequence}(b,c)$, indicating that both the occurrences of activity $b$ after $a$ and the occurrences of activity $c$ after $b$ meet some support threshold. The binary relations $\mathit{sequence}(a,b)$ and $\mathit{sequence}(b,c)$ combined do not imply a tertiary relation equivalent to process tree $\rightarrow(a,\rightarrow(b,c))$, since it could be the case that specifically those occurrences of $b$ that are preceded by $a$ are rarely followed by $c$. The LPM discovery approach discussed in this paper enables discovery of relations between three or more activities.

Hybrid process discovery \cite{Maggi2014} aims at discovering a process model that consists partially of procedural process model constructs and partially of declarative process model constructs. Existing hybrid process discovery approaches consist of a first step where activities are separated into a group of structured activities and a group of unstructured activities, based on the number of unique predecessors and successors of an activity. However, some activities, such as activity $a$ in the event log of Figure \ref{sfig:traces_example}, are part of a frequent pattern, but also to occur as noise at random point in the traces. Such activities would be classified as noisy activities by existing hybrid process discovery approaches, resulting in this activity being modeled with binary declarative constructs.

The Fuzzy Miner \cite{Gunther2007} is a process discovery technique developed to deal with complex and flexible process models. It connects nodes that represent activities with edges indicating follows relations, taking into account the relative significance of follows/precedes relations and allowing the user to filter out edges using a slider. However, the process models obtained using the Fuzzy Miner lack formal semantics, e.g. when a node has two or more outgoing edges, it is undefined whether this represents a choice, an exclusive choice, or parallel execution of the connected nodes.\looseness=-1

We have described several techniques that are related in that sense that i) they aim to enable mining of process insight from less structured processes on which traditional process discovery methods fail, or ii) they aim to extract a collection of process models that each represent some subprocess. However, none of the existing techniques in category (i) is able to deal with event logs where some frequent patterns are surrounded by random events, as is the case in the event log in Figure \ref{sfig:traces_example}, and all of the existing methods in category (ii) either require a completely different type of input (a collection of graphs), or they support only a part of the constructs supported by the LPM discovery approach.

\section{Local Process Model Discovery Approach}
\label{sec:approach}
A local process model (LPM) aims to describe frequent behavior in an event log in local, i.e. smaller, patterns, typically between three and five activity nodes. A LPM does not aim to describe all behavior in a trace completely, but instead aims to describe traces partially, focusing on frequently occurring patterns. A LPM $\LN$ represents behavior over $\Sigma_M$ and accepts language $\Lan(LN)$. The closure of the accepting language with respect to alphabet $\Sigma_L$ is defined as $\overline{\Lan}(\LN, \Sigma_L)=\{\sigma\in{\Sigma_L}^*|\;\sigma\upharpoonright_{\Sigma_M}\in\Lan(\LN)\}$.\\

\begin{figure}[t]
	\centering
	\subfloat{
		\begin{tikzpicture}
		\begin{scope}[thick,scale=0.6, level distance=1cm, every node/.style={scale=0.8},level 1/.style={sibling distance=20mm},level 2/.style={sibling distance=10mm}]
		\node(5)[sibling distance=16mm]{} 
		child{node(Q){a}};
		\end{scope}
		\end{tikzpicture}
	}
	\qquad
	\subfloat{
		\begin{tikzpicture}
		\begin{scope}[thick,scale=0.6, level distance=1cm, every node/.style={scale=0.8},level 1/.style={sibling distance=20mm},level 2/.style={sibling distance=10mm}]
		\node(5)[sibling distance=16mm]{} 
		child{node(Q){b}};
		\end{scope}
		\end{tikzpicture}
	}
	\qquad
	\raisebox{0.7\height}{
		\subfloat{$\cdots$}
	}
	\qquad
	\subfloat{
		\begin{tikzpicture}
		\begin{scope}[thick,scale=0.6, level distance=1cm,every node/.style={scale=0.8},level 1/.style={sibling distance=20mm},level 2/.style={sibling distance=10mm}]
		\node(5)[sibling distance=16mm]{} 
		child{node(Q){z}};
		\end{scope}
		\end{tikzpicture}
	}
	\caption{Set of elementary process trees over $\Sigma_L=\{a,b,\dots,z\}$.}
	\label{fig:elementary_tree_set}
\end{figure}
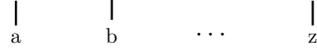

\noindent Here, we introduce a local process model discovery approach that consists of four main steps:
\begin{description}
	\item[1) Generation]{Generate the initial set $CM_1$ (so $i=1$) of candidate LPM in the form of process trees consisting of one leaf for each activity $a\in\Sigma_L$. Figure \ref{fig:elementary_tree_set} shows this set of elementary process trees for an event log over alphabet $\Sigma_L=\{a,b,\dots,z\}$. Create selected set of selected LPMs $SM=\emptyset$.}
	\item[2) Evaluation]{Evaluate LPMs in current candidate set $CM_i$ based on a set of quality criteria.}
	\item[3) Selection]{Based on evaluation results, a set $SCM_i\subseteq CM_i$ of candidate LPMs are selected. $SM=SM\cup SCM_i$. If $SCM_i=\emptyset$ or $i\ge max\_iterations$: stop.}
	\item[4) Expansion]{Expand $SCM_i$ into a set of larger, expanded, candidate process models, $CM_{i+1}$. Goto step 2 using the newly created candidate set $CM_{i+1}$.}
\end{description}


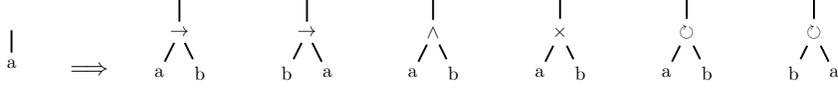
\begin{figure}[t]
	\vspace{-0.5cm}
	\centering
	\scalebox{0.9}{
	\raisebox{0.2\height}{
		\subfloat{
			\begin{tikzpicture}
			\begin{scope}[thick,scale=0.6,  level distance=1cm, every node/.style={scale=0.8},level 1/.style={sibling distance=20mm},level 2/.style={sibling distance=10mm}]
			\node(5)[sibling distance=16mm]{} 
			child{node(Q){a}};
			\end{scope}
			\end{tikzpicture}
		}}
		\quad
		\raisebox{1.25\height}{
			\subfloat{$\Longrightarrow$}
		}
		\quad
		\subfloat{
			\begin{tikzpicture}
			\begin{scope}[thick,scale=0.6,  level distance=1cm, every node/.style={scale=0.8},level 1/.style={sibling distance=20mm},level 2/.style={sibling distance=10mm}]
			\node(5)[sibling distance=16mm]{} 
			child{
				node(Q){$\rightarrow$}{
					child{node(W){a}}
					child{node(E){b}}
				}
			};
			\end{scope}
			\end{tikzpicture}
		}
		\qquad
		\subfloat{
			\begin{tikzpicture}
			\begin{scope}[thick,scale=0.6,  level distance=1cm, every node/.style={scale=0.8},level 1/.style={sibling distance=20mm},level 2/.style={sibling distance=10mm}]
			\node(5)[sibling distance=16mm]{} 
			child{
				node(Q){$\rightarrow$}{
					child{node(W){b}}
					child{node(E){a}}
				}
			};
			\end{scope}
			\end{tikzpicture}
		}
		\qquad
		\subfloat{
			\begin{tikzpicture}
			\begin{scope}[thick,scale=0.6,  level distance=1cm, every node/.style={scale=0.8},level 1/.style={sibling distance=20mm},level 2/.style={sibling distance=10mm}]
			\node(5)[sibling distance=16mm]{} 
			child{
				node(Q){$\wedge$}{
					child{node(W){a}}
					child{node(E){b}}
				}
			};
			\end{scope}
			\end{tikzpicture}
		}
		\qquad
		\subfloat{
			\begin{tikzpicture}
			\begin{scope}[thick,scale=0.6,  level distance=1cm, every node/.style={scale=0.8},level 1/.style={sibling distance=20mm},level 2/.style={sibling distance=10mm}]
			\node(5)[sibling distance=16mm]{} 
			child{
				node(Q){$\times$}{
					child{node(W){a}}
					child{node(E){b}}
				}
			};
			\end{scope}
			\end{tikzpicture}
		}
		\qquad
		\subfloat{
			\begin{tikzpicture}
			\begin{scope}[thick,scale=0.6, level distance=1cm, every node/.style={scale=0.8},level 1/.style={sibling distance=20mm},level 2/.style={sibling distance=10mm}]
			\node(5)[sibling distance=16mm]{} 
			child{
				node(Q){$\circlearrowright$}{
					child{node(W){a}}
					child{node(E){b}}
				}
			};
			\end{scope}
			\end{tikzpicture}
		}
		\qquad	
		\subfloat{
			\begin{tikzpicture}
			\begin{scope}[thick,scale=0.6, level distance=1cm, every node/.style={scale=0.8},level 1/.style={sibling distance=20mm},level 2/.style={sibling distance=10mm}]
			\node(5)[sibling distance=16mm]{} 
			child{
				node(Q){$\circlearrowright$}{
					child{node(W){b}}
					child{node(E){a}}
				}
			};
			\end{scope}
			\end{tikzpicture}
		}}
		\caption{The set of process tree expansion operations to leaf node $a$, where $b\in\Sigma_L$.}
		\label{fig:tree_growth}
	\end{figure}

Expansion consists of the replacement of one of the leaf activity nodes $a$ of the process tree by an operator node of one of the operator types, where one of the child nodes is the replaced activity node $a$ and the other is a new activity node representing one of the activities $b\in\Sigma_L$. Figure \ref{fig:tree_growth} shows the set of expansion operations to leaf node $a$, consisting of six possible expansions. Two types of expansion operations are defined for the sequence ($\rightarrow$) and the loop ($\circlearrowright$) operator types, as ${\rightarrow(a,b)}\;{\not\equiv}\;{\rightarrow(b,a)}$ and ${\circlearrowright(a,b)}\;{\not\equiv}\;{\circlearrowright(b,a)}$. However, only one expansion operation is needed for the $\wedge$ and $\times$ operators because of their symmetry (${\wedge(a,b)}\;{\equiv}\;{\wedge(b,a)}$ and ${\times(a,b)}\equiv{\times(b,a)}$). We call $\wedge$ and $\times$ symmetrical operators.

\begin{wrapfigure}{r}{0.25\textwidth}
	\vspace{-0.5cm}
	\begin{tikzpicture}
		\begin{scope}[thick,scale=0.6, level distance=1cm, every node/.style={scale=0.8},level 1/.style={sibling distance=20mm},level 2/.style={sibling distance=10mm}]
		\node(5)[sibling distance=16mm]{} 
		child{node(Q){$\times$} 
			child{node(W) {$\times$}
				child{node(E){$a$}}
				child{node(R){$c$}}
			}
			child{node(T) {$b$}}
		};
		\end{scope}
	\end{tikzpicture}
	\raisebox{1cm}{
	$\equiv$}
	\begin{tikzpicture}
		\begin{scope}[thick,scale=0.6, level distance=1cm, every node/.style={scale=0.8},level 1/.style={sibling distance=20mm},level 2/.style={sibling distance=10mm}]
		\node(5)[sibling distance=16mm]{} 
		child{node(Q){$\times$} 
			child{node(T) {$a$}}
			child{node(W) {$\times$}
				child{node(E){$b$}}
				child{node(R){$c$}}
			}
		};
		\end{scope}
	\end{tikzpicture}
\end{wrapfigure}
Expanding the first leaf node ($a$) of process tree ${\times(a,b)}$ with the $\times$ operator and some activity $c\in\Sigma_L$ results in the leftmost process tree depicted on the right, while applying the same expansion the second leaf node ($b$) of the same process tree results in a behaviorally equivalent process tree (as shown on the right). The same holds for expansion of ${\wedge(a,b)}$ with the $\wedge$ operator. Therefore, the number of expansions can be reduced further by restricting expansion of a leaf node that has a symmetrical operator as parent with the same symmetrical operator only to the rightmost child. This prevents unnecessary computation by generating both of the behaviorally equivalent trees shown on the right.

The number of possible expansion operations for a process tree $P$ grows with the size of the alphabet of the event log $|\Sigma_L|$ and the number of activity nodes in $P$. This is easy to see, as each type of expansion operation can be applied to each activity node in $P$, leading to $6\times|\Sigma_L|$ expansion operations per activity node. At every point in the expansion process, the number of activity nodes in the tree is equal to the number of expansion operations performed plus one, as each expansion operation adds one activity node to the process tree.

The local process model discovery procedure stops when no process tree in the current set of candidate process models meets the quality criteria, or, to guarantee termination, when a maximum number of expansion steps, \\$max\_iterations$, is reached.

The approach of iteratively expanding, selecting, and expanding process trees described above is not theoretically limited to the set of operator nodes described above, and can easily be extended to other operators, such as an inclusive choice operator or a long-term dependency operator. Adding extra operators, however, comes with the price of increased computational complexity as it increases the number of ways to expand a process tree.

	\section{Quality Criteria \& Metrics}
	\label{sec:quality}
	Assume for now that we have an oracle function $\lambda_\LN$ that generates a segmentation of a given trace $\sigma$: $\lambda_\LN(\sigma)=\gamma_0\xi_1\gamma_1\xi_2\dots\xi_k\gamma_k$, with $\xi_i\in\Lan(\LN)$ and $\gamma_i\notin\Lan(\LN)$, such that the number of events in $\{\xi_1,\dots,\xi_k\}$ is maximized: the higher the number of events in $\xi_i\in\Lan(\LN)$ segments, the larger the share of trace $\sigma$ explained by the LPM. A higher value $k$ indicates more frequent presence of LPM execution traces in $\sigma$. $\lambda_\LN^k(\sigma)$ denotes the number of $\xi\in\Lan(\LN)$ segments in $\lambda_\LN(\sigma)$. $\lambda_\LN^\xi(\sigma)$ denotes the multiset of segments $\xi_i\in\Lan(\LN)$. Here we discuss a number of quality criteria for LPM with regard to an event log.\\
	
	\noindent\textbf{Support}
	Relates to the number of fragments in the event log that can be considered to be an instance of the LPM under evaluation. The rationale behind this quality dimension: an LPM whose execution traces are observed more frequently in the event log represents it better. We transform the count of pattern instances of $\LN$ in $L$ into a $[0,1)$-interval number through the following transformation:\looseness=-1
	
	$\mathit{support}(\LN,L)=\frac{\sum_{\sigma\in L}\lambda_\LN^k(\sigma)}{(\sum_{\sigma\in L}\lambda_\LN^k(\sigma))+1}$\\
	
	\noindent\textbf{Confidence}
	An event fits an LPM when it is part of a segment $\xi\in\Lan(LN)$. The confidence of event type $e\in\Sigma_M$ in LPM $\LN$ given event log $L$, is the ratio of events of type $a$ in $L$ that fit $\LN$:
	
	$\mathit{confidence}(a,L) = \frac{\sum_{\sigma\in L}\#_a(\lambda_{\LN}^\xi(\sigma))}{\#_a(L)}$\\
	We use the harmonic mean to aggregate confidence values for individual activities to a single metric, as it is more sensitive to a single lower-than-average value than the geometric mean. We define the confidence of an LPM $\LN$ given an event log $L$ to be the harmonic mean of the individual confidence scores of the event types of $\LN$:
	
	$\mathit{confidence}(\LN,L)=\frac{|\Sigma_M|}{\sum_{a\in \Sigma_M} \frac{1}{\mathit{confidence}(a,L)}}$\\
	
	\noindent\textbf{Language Fit}
	Language fit expresses the ratio of the behavior allowed by the LPM that is observed in the event log. LPMs that allow for much more behavior than what is observed are likely to overgeneralize and therefore do not describe the behavior in the event log well. The language fit of an LPM $\LN$ given log $L$ is:
	
	$\mathit{language\_fit}(\LN,L)=\frac{|\{\phi\in\Lan(\LN)|\exists\sigma\in L:\phi\in\lambda^\xi_\LN(\sigma)\}|}{|\Lan(\LN)|}$\\
	
	\noindent Since $|\Lan(\LN)|=\infty$ in case $\LN$ contains a loop, $language\_fit(\LN,L)=0$ for any $\LN$ containing a loop. Restricting the language and the LPM instances to sequences of a bounded length allows us to approximate language fit for models with infinite size language. Language fit restricted to bound $n\in\mathbb{N}$ is defined as:
	
	$\mathit{language\_fit_n}(\LN,L)=\frac{|\{\phi\in\Lan_n(\LN)|\exists\sigma\in L:\phi\in\lambda^\xi_\LN(\sigma)\}|}{|\Lan_n(\LN)|}$\\

	\begin{wrapfigure}{r}{0.15\textwidth}
		\vspace{-16pt}
		\begin{center}
			\begin{tikzpicture}
			\begin{scope}[thick,scale=0.6,  level distance=1cm, every node/.style={scale=0.8},level 1/.style={sibling distance=20mm},level 2/.style={sibling distance=10mm}]
			\node(5)[sibling distance=16mm]{} 
			child{
				node(Q){$\times$}{
					child{node(W){a}}
					child{node(E){$\times$}{
							child{node(R){b}}
							child{node(T){$\times$}{
									child{node(Y){c}}
									child{node(U){d}}
								}
							}
						}
					}
				}
			};
			\end{scope}
			\end{tikzpicture}
		\end{center}
		\vspace{-20pt}
	\end{wrapfigure}
	\noindent\textbf{Determinism}
	Flower-like process trees, like the one shown on the right, are not desirable as they provide little insight in what behavior we are likely to observe. Deterministic LPMs have more predictive value in with regard to future behavior.
	When the language of LPM $\LN$ contains traces if type $\sigma a \gamma_1$ and $\sigma b \gamma_2$, the continuation of the trace after observing prefix $\sigma$ can be either $a$ or $b$, leaving some uncertainty. LPMs with a high degree of certainty are preferable over LPMs with a low degree of certainty. A metric for the determinism quality dimension is dependent on the process model and not only on its language. Let $\mathcal{R}(\LN)$ be a set of reachable states of an LPM $\LN$. $\mathcal{W}_L:\mathcal{R}(\LN)\to\mathbb{N}$ represents a function assigning the number of times a state is reached while replaying the fitting segments of log $L$ on $\LN$. $\mathcal{D}:\mathcal{R}(\LN)\to\mathbb{N}$ represents a function assigning the number of transitions enabled in a certain state in $\LN$. Determinism is defined as:\\
	
	$\mathit{determinism}(\LN,L)=\frac{\sum_{m\in\mathcal{R}(\LN)} \mathcal{W}_L(m)}{\sum_{m\in\mathcal{R}(\LN)} \mathcal{W}_L(m)\cdot \mathcal{D}(m)}$\\
	
	\noindent\textbf{Coverage}
	Let $\LN$ be an LPM and $L$ be an event log. Let $\#_*(L)$ denote the total number of events of event log $L$. Coverage is defined as the ratio of the number of events in the log after projecting the event log on the labels of $\LN$ to the number of all events in the log:
		
	$\mathit{coverage}(\LN,L) = \frac{\#_*(L\upharpoonright_{\Sigma_M})}{\#_*(L)}$
	
	\subsection{Local Process Model Selection \& Ranking}
	The quality dimensions and metrics defined are used to select and rank local process models generated through the recursive process tree exploration approach. Often, one is interested in multiple quality criteria at the same time. A high-support local process model that has a low determinism score (e.g., a small flower pattern) does not generate much process insight, while a deterministic pattern that has low support does not describe the behavior in the log very well. So it is possible to set thresholds per dimension. It is also useful to rank patterns according to a weighted average over the quality criteria. The appropriate weighting of the quality dimensions depends on the business questions and the situation at hand.
	
	\subsection{Monotonicity Properties \& Pruning}
	\label{sec:monotonicity}
	Often one is not interested in local process models with a low support, confidence, or determinism. Setting a minimum threshold for these quality criteria allows us to prune away those parts of the search space where we know that expansions of a candidate local process model can never meet the criteria because of monotonicity, resulting in a speedup of the proposed recursive process tree exploration procedure. Pruning based on monotonicity is similar to the pruning performed in the well-known Apriori algorithm \cite{Agrawal1994}, and other algorithms inspired by the Apriori algorithm, such as \cite{Agrawal1995}.
	
	Any expansion of process tree $P$ where a leaf node $a\in P$ is replaced by subtree ${\rightarrow(a,b)}$, ${\rightarrow(b,a)}$, ${\wedge(a,b)}$, or ${\circlearrowright(a,b)}$ for any $b\in\Sigma_L$ is guaranteed to be less frequent, i.e. has lower support, than $P$. The intuition behind this is that expansion put additional requirements of the alignments, possibly causing some fitting segments for a trace $\sigma$ obtained by $\lambda_P^\xi(\sigma)$ to not fit the expansion of $P$. Therefore, when $P$ does not meet support threshold $min_{sup}$, its expansions of any activity node $a$ into $\rightarrow(a,b)$, $\rightarrow(b,a)$, $\wedge(a,b)$, and $\circlearrowright(a,b)$ can be pruned from the search space.
	
	Process tree $P$ is guaranteed to be at least as deterministic as its expansion where activity node $a\in P$ is replaced by subtree $\times(a,b)$ or $\wedge(a,b)$ for any $b\in\Sigma_L$. Therefore, when $P$ does not meet determinism threshold $min_{det}$, its expansions of any activity node $a$ into $\times(a,b)$, and $\wedge(a,b)$ can be pruned from the search space.\looseness=-1

	\section{Alignment-Based Evaluation of Local Process Models}
	\label{sec:alignment_based_evaluation}
	We now describe a way to define function $\lambda_\LN$. We evaluate LPMs using Petri nets because of the rich set of analysis techniques available for Petri nets. Important for the definition of $\lambda_\LN$ is the notion of alignments \cite{Aalst2012}, which aims to find a sequence of model firings starting at the initial marking and ending in a final marking that is an optimal approximation of the behavior in the event log. Alignments relate model traces and event log traces through a series of three types of moves: \emph{synchronous moves}, \emph{moves on model}, and \emph{moves on log}. When an event in the event log trace can be performed in the process model, log and model can move \emph{synchronously}. However, when a trace of the log does not fit the model, log and model cannot move synchronously from the start to the end of the trace. A \emph{move on model} corresponds to a firing of a transition in the model that cannot be mapped to an event in the log. A \emph{move on log} corresponds to an event in the log that cannot be mapped to a transition firing in the model. Since both \emph{moves on model} and \emph{moves on log} are suboptimal behavior, they are often assigned certain costs such that the alignment will only chose to do \emph{moves on model} or \emph{moves on log} when these moves are unavoidable. \emph{Moves on model} enable the unwanted behavior that a partial execution of the LPM can be identified as an LPM execution trace. To avoid this behavior, we use a version of alignments where moves on model on non-silent transitions are prohibited (infinite costs).
	
\begin{figure*}[t]
		\centering
		\subfloat[]{
			\centering
			\scalebox{0.7}{
				\begin{tikzpicture}
				[node distance=1.2cm,
				on grid,>=stealth',
				bend angle=20,
				auto,
				every place/.style= {minimum size=5mm},
				every transition/.style = {minimum size = 5mm},
				transitionH/.style={rectangle, thick, fill=black, minimum width=3mm, inner ysep=9pt }
				]
				\node [place, tokens = 1] (p){};
				\node [transition] (2) [label=below:$t_1$,align=center, right = of p] {A}
				edge [pre] node[auto] {} (p);
				\node [place] (p3) [above right = of 2] {}
				edge[pre] node[auto] {} (2);
				\node [transition] (t1) [label=below:$t_2$,right = of p3]{B}
				edge[pre] node[auto] {} (p3);
				\node [place] (p4) [below right = of 2] {}
				edge[pre] node[auto] {} (2);
				\node [transition] (t2) [label=below:$t_3$,right = of p4] {C}
				edge[pre] node[auto] {} (p4);
				\node [place] (p5) [right = of t1] {}
				edge[pre] node[auto] {} (t1);
				\node [place] (p6) [right = of t2] {}
				edge[pre] node[auto] {} (t2);
				\node [transitionH] (t3) [label=below:$t_4$,below right = of p5] {}
				edge[pre] node[auto] {} (p5)
				edge[pre] node[auto] {} (p6);
				\node [place,pattern=custom north west lines,hatchspread=1.5pt,hatchthickness=0.25pt,hatchcolor=gray] (p7) [right=of t3] {}
				edge[pre] node[auto] {}(t3);
				\end{tikzpicture}
			}
			\label{sfig:before:backloop}
		}
		\subfloat[]{
			\centering
			\scalebox{0.7}{
				\begin{tikzpicture}
				[node distance=1.2cm,
				on grid,>=stealth',
				bend angle=20,
				auto,
				every place/.style= {minimum size=5mm},
				every transition/.style = {minimum size = 5mm},
				transitionH/.style={rectangle, thick, fill=black, minimum width=3mm, inner ysep=9pt }
				]
				\node [place, tokens = 1, pattern=custom north west lines,hatchspread=1.5pt,hatchthickness=0.25pt,hatchcolor=gray] (p){};
				\node [transition] (2) [label=below:$t_1$,align=center, right = of p] {A}
				edge [pre] node[auto] {} (p);
				\node [place] (p3) [above right = of 2] {}
				edge[pre] node[auto] {} (2);
				\node [transition] (t1) [label=below:$t_2$,right = of p3]{B}
				edge[pre] node[auto] {} (p3);
				\node [place] (p4) [below right = of 2] {}
				edge[pre] node[auto] {} (2);
				\node [transition] (t2) [label=below:$t_3$,right = of p4] {C}
				edge[pre] node[auto] {} (p4);
				\node [place] (p5) [right = of t1] {}
				edge[pre] node[auto] {} (t1);
				\node [place] (p6) [right = of t2] {}
				edge[pre] node[auto] {} (t2);
				\node [transitionH] (t3) [label=below:$t_4$,below right = of p5] {}
				edge[pre] node[auto] {} (p5)
				edge[pre] node[auto] {} (p6);
				\node [place] (p7) [right=of t3] {}
				edge[pre] node[auto] {}(t3);
				\node [transitionH] (p8) [label=below:$t_{bl_1}$,above=of t1]{}
				edge[bend left,pre] node[auto] {}(p7)
				edge[bend right, post] node[auto]
				{}(p);
				\end{tikzpicture}
				
			}
			\label{sfig:after_backloop}
		}
		\\
		\subfloat[]{
			\centering
			\scalebox{0.9}{
				\begin{tabular}{|l|l|l|l|l|l|l|l|l|l|l|l|l|l|l|}
					A&A&C&B&$\gg$&$\gg$&A&A&C&B&$\gg$&$\gg$&B&C\\
					\hline
					A&$\gg$&C&B&$\tau$&$\tau$&A&$\gg$&C&B&$\tau$&$\tau$&$\gg$&$\gg$\\
					\hline
					$t_1$&&$t_3$&$t_2$&$t_4$&$t_{bl_1}$&$t_1$&&$t_3$&$t_2$&$_4$&$t_{bl_1}$&&\\
				\end{tabular}}
				\label{sfig:alignment}
			}
			\caption{\emph{(a)} An example local process model, $\LN$. \emph{(b)} Process model $\LN_{BL}$, constructed from $\LN$ by adding a silent connection from final to initial marking, and the final marking set to the initial marking. \emph{(c)} Alignment of the trace $t_1$ on $\LN_{BL}$ when disallowing model moves on non-silent transitions.}
			\label{fig:backloop_connection_example}
			\vspace{-0.1cm}
		\end{figure*}
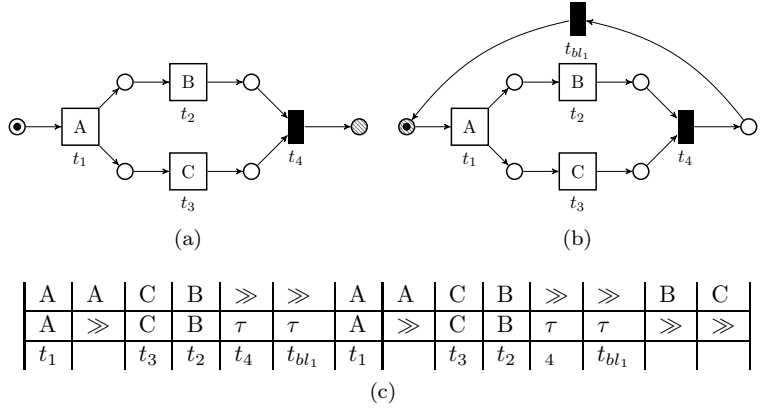

	Alignments aim to match an event log trace with a single execution of a process model. However, an event log trace can contain more than one execution trace of an LPM. We modify the Petri net representation of the LPM such that we connect each final marking to the initial marking through a silent transition, allowing the alignment to relate a single trace to multiple executions of the model. Figure \ref{sfig:before:backloop} shows an example LPM and Figure \ref{sfig:after_backloop} shows the corresponding Petri net after transformation. We transform LPM $\LN(N,M_0,\MF)$ with $N=(P,T,F,\Sigma_M,\ell)$ into $\LN_{BL}(N_{BL},M_0,\{M_0\})$ with $N_{BL}=(P,T_{BL},F_{BL},\Sigma_M,\ell_{BL})$, such that:
	\begin{itemize}
		\item $T_{BL} = T \cup \{t_{bl_M}|M\in\MF\}$,
		\item $F_{BL} = F \cup \{(p,t_{bl_M})|M\in\MF\wedge p\in M\} \cup \{t_{bl_M}|M\in\MF\wedge p\in M_0\}$,
		\item $\ell_{BL} \in T_{BL} \rightarrow \Sigma_M \cup \{\tau\}$ with:\\
		$\ell_{BL} =
		\begin{cases}
		\ell(T),& \text{if } t\in T,\\
		\tau,              & \text{otherwise.}
		\end{cases}$
	\end{itemize}
	
		$\LN_{BL}$ contains a set of added silent transitions, $\{t_{bl_M}|M\in\MF\}$, consisting of one silent transition for each final marking $M \in \MF$. $Backloop:\MF\to T_{bl}$ is a bijective mapping from a final marking $M\in \MF$ to a silent transition $t_{bl_M}$. A silent transition $t_{bl_M}$ has all places in final marking $M$ as input and place $M_0$ as output. The number of executions of backloop transitions $\{t_{bl_M}|M\in\MF\}$ in the alignments of $L$ on $\LN$ indicates the number of executions of traces of $\LN$ in $L$. Note that alignments require the model to be in a marking $M\in \MF$ at the end of the alignment. This is impossible to obtain when pattern $\LN$ is absent in log $L$. Therefore, we set the final marking to $\{M_{0}\}$, allowing the alignments to make a complete sequence of \emph{moves on log}, resulting in zero executions of the model.\looseness=-1

		Figure \ref{sfig:alignment} illustrates the concept of alignments through an example, showing the alignment of the non-fitting trace $\langle A,A,C,B,A,A,C,B,B,C\rangle$ on the model of Figure \ref{sfig:after_backloop}. The top row of the alignments represents the behavior of the log, while the middle row and the bottom row represent the behavior of the model. $\gg$ indicates \emph{no move}, with a $\gg$ in the top row indicating a \emph{move on model} and in the middle row indicating a \emph{move on log}. The model is able to mimic the first event of the trace by executing $t_1$ with label $A$, but is not able to mimic the second $A$ in the log, resulting in a \emph{move on log}. The $C$ and $B$ in the log can be mimicked (by $t_3$ and $t_2$ respectively). Next event $A$ in the log can only be mimicked by the model by first firing $t_{bl_1}$, resulting in a \emph{move on model}, represented by the $\gg$ in the log. Afterwards, $A$ can be mimicked and another \emph{move on log} is needed for the second $A$. $C$ and $B$ can again be mimicked, after which a \emph{move on log} is again needed as the log cannot mimic $t_{bl_1}$. Would we not have prohibited \emph{moves on models} on non-silent transition, the alignment could now have executed a \emph{move on model} on $A$, enabling synchronous moves on both $B$ and $C$, falsely giving the impression that the LPM would have a third occurrence in the trace. As we prohibited the \emph{model move} on $A$, the only option is to decide a \emph{move on log} on $B$ and $C$, thereby not counting the incomplete occurrence of the pattern.
		
		LPM $\LN$ is evaluated on event log $L$ by projecting $L$ on the set of labels of $\LN$, $L'=L\upharpoonright_{\Sigma_M}$. The middle row of the alignment of $L'$ on $\LN_{BL}$ represents the segmentation $\lambda_\LN^\xi$, where $\tau$ moves on a transition ${t_{bl}}_i\in\{{{t_{bl}}_m}|M\in MF\}$ indicates the start of a new segment. The alignment in Figure \ref{sfig:alignment} shows that $\lambda_\LN^\xi(\langle A,A,C,B,A,A,C,B,B,C \rangle)=[\langle A,C,B \rangle^2]$.

		\label{ssec:determinism_petri}
		\begin{table}[t]
			\centering
			\vspace{0.2cm}
			\begin{tabular}{|l|l|l|l|l|l|l|l|l|l|l|}
				A&C&B&$\tau$&$\tau$&A&C&B&$\tau$&$\tau$\\
				\hline
				$t_1$&$t_3$&$t_2$&$t_4$&$t_{bl_1}$&$t_1$&$t_3$&$t_2$&$_4$&$t_{bl_1}$\\
				\hline
				1&2&1&1&1&1&2&1&1&1\\
			\end{tabular}
			\caption{Number of transitions enabled at each point in the alignment}
			\label{tab:det_align}
			\vspace{-0.5cm}
		\end{table}
		\subsection{Determinism on Petri nets}
		We now explain through an example how to calculate determinism for Petri nets. Each transition firing in a Petri net corresponds to a change in the marking of the net. Table  \ref{tab:det_align} shows the transitions fired in the alignment of Figure \ref{sfig:alignment}. The bottom row represents the number of transitions that were enabled in the Petri net when the transition fired. When $t_3$ fired, the Petri net was in a marking where both $t_2$ and $t_3$ were enabled. The determinism of the net corresponds to one divided by the average number of enabled transitions during replay. In the example, $\mathit{determinism}(\LN,L)=\frac{10}{12}$.
		
		\section{Case Studies}
		\label{sec:case_studies}
		We now evaluate the proposed local process model mining method on two real life data sets.
\subsection{BPIC '12 Data Set}
The Business Process Intelligence Challenge (BPIC)'12 data set originates from a personal loan or overdraft application process in a global financial institution. We transformed the event log to obtain traces of all activities in a single day performed by one specific resource (bank employee). This resource was selected randomly to be resource id $10939$. The event log for this specific resource contains 49 cases (working days), 2763 events, and 14 activities. Discovering the local process models with the approach described in this paper took 34 seconds on a machine with a 4-core 2.4 GHz processor using a support threshold of 0.7.

\begin{figure}[t]
	\centering
	\includegraphics[width=0.87\textwidth]{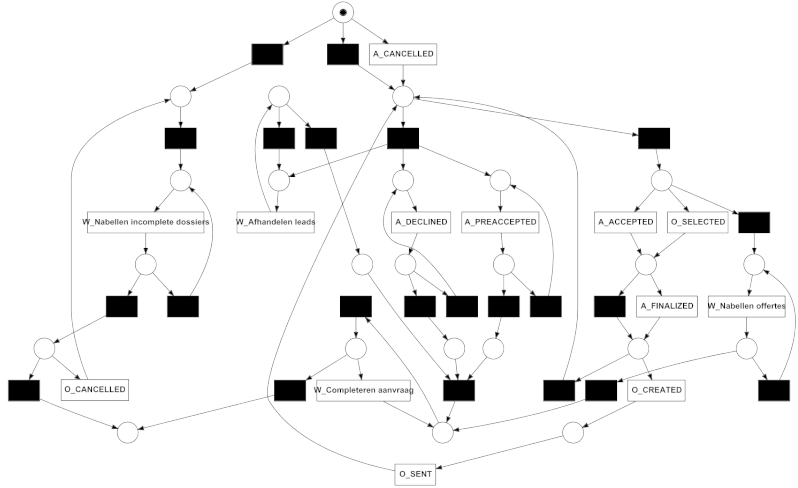}
	\caption{Process model of the behavior of resource $10939$ in the BPIC'12 log, obtained using the Inductive Miner infrequent (20\%).}
	\label{fig:bpic_inductive}
	\vspace{-0.5cm}
\end{figure}
\begin{figure}[!htb]
	\captionsetup[subfigure]{labelformat=empty}
	\centering
	\subfloat{
		\raggedleft
		\subfloat{
			\hspace{1.16cm}
			\raggedleft
			\raisebox{0.9\height}{(a)}
			\scalebox{0.75}{
				\begin{tikzpicture}
				[node distance=1.4cm,
				on grid,>=stealth',
				bend angle=20,
				auto,
				every place/.style= {minimum size=1mm},
						every transition/.style = {minimum size = 3mm, text width=1.5cm,align=center,font=\scriptsize}
				]
				\node [place, tokens = 1] (p){};
				\node [transition] (2) [right = of p] {\textbf{O\_SELEC- TED} 124/124}
				edge [pre] node[auto] {} (p);
				\node [place] (p3) [right = of 2] {}
				edge[pre] node[auto] {} (2);
				\node [transition] (t1) [right = of p3]{\textbf{O\_CREA- TED} 124/124}
				edge[pre] node[auto] {} (p3);
				\node [place] (p5) [right = of t1] {}
				edge[pre] node[auto] {} (t1);
				\node [transition] (t3) [right = of p5] {\textbf{O\_SENT} 124/124}
				edge[pre] node[auto] {} (p5);
				\node [place,pattern=custom north west lines,hatchspread=1.5pt,hatchthickness=0.25pt,hatchcolor=gray] (p7) [right=of t3] {}
				edge[pre] node[auto] {}(t3);
				\end{tikzpicture}}}
		\subfloat{
			\raggedleft
			\raisebox{.57\height}{
				\scalebox{0.7}{
					\begin{tabular}{|l|l|}
						\hline
						Support & 0.9920\\
						Confidence& 1.0000\\
						Language fit$_5$& 1.0000\\
						Determinism& 1.0000\\
						Coverage& 0.1346\\
						\hline
					\end{tabular}}}
				}
			}
			\\
			\centering
			\vspace{-0.5cm}
			\subfloat{
				\raggedleft
				\subfloat{
					\hspace{-0.9cm}
					\raggedleft
					\raisebox{0.9\height}{(b)}
					\scalebox{0.75}{
						\begin{tikzpicture}
						[node distance=1.4cm,
						on grid,>=stealth',
						bend angle=20,
						auto,
						every place/.style= {minimum size=1mm},
						every transition/.style = {minimum size = 3mm, text width=1.5cm,align=center,font=\scriptsize}
						]
						\node [place, tokens = 1] (p){};
						\node [transition] (2) [right = of p] {\textbf{A\_ACCE- PTED} 103/104}
						edge [pre] node[auto] {} (p);
						\node [place] (p3) [right = of 2] {}
						edge[pre] node[auto] {} (2);
						\node [transition] (t1) [right = of p3]{\textbf{O\_SELEC- TED} 103/124}
						edge[pre] node[auto] {} (p3);
						\node [place] (p5) [right = of t1] {}
						edge[pre] node[auto] {} (t1);
						\node [transition] (t3) [right = of p5] {\textbf{O\_CREA- TED} 103/124}
						edge[pre] node[auto] {} (p5);
						\node [place] (p6) [right = of t3] {}
						edge[pre] node[auto] {} (t3);
						\node [transition] (t4) [right = of p6] {\textbf{O\_SENT} 103/124}
						edge[pre] node[auto] {} (p6);
						\node [place,pattern=custom north west lines,hatchspread=1.5pt,hatchthickness=0.25pt,hatchcolor=gray] (p7) [right = of t4] {}
						edge[pre] node[auto] {} (t4);
						\end{tikzpicture}}

				}
				\subfloat{
					\hspace{-0.22cm}
					\raggedleft
					\raisebox{.55\height}{
						\scalebox{0.7}{
							\begin{tabular}{|l|l|}
								\hline
								Support & 0.9904\\
								Confidence& 0.8655\\
								Language fit$_5$& 1.0000\\
								Determinism& 1.0000\\
								Coverage& 0.1723\\
								\hline
							\end{tabular}
						}}
					}
				}
				\\
				\vspace{-.5cm}
				\subfloat{
					\centering
					\subfloat{
						\raggedleft
						\hspace{-0.21cm}
						\raisebox{2.9\height}{(c)}
						\scalebox{0.75}{
							\begin{tikzpicture}
							[node distance=1.4cm,
							on grid,>=stealth',
							bend angle=20,
							auto,
							every place/.style= {minimum size=0.1mm},
							every transition/.style = {minimum size = 3mm, text width=1.5cm,align=center,font=\scriptsize},
							transitionH/.style={rectangle, thick, fill=black, minimum width=3mm, inner ysep=9pt }
							]
							\node [place, tokens = 1] (p){};
							\node [transitionH] (sil) [right = of p] {}
							edge[pre] node[auto] {} (p);
							
							\node [place] (p1) [above right = 0.7 and 1 cm of sil] {}
							edge [pre] node[auto] {} (sil);
							\node [transition] (t1) [right = of p1] {\textbf{O\_SELEC- TED} 104/124}
							edge [pre] node[auto] {} (p1);
							\node [place] (p2) [ right = of t1] {}
							edge [pre] node[auto] {} (t1);
							
							\node [place] (p3) [below right = 0.7 and 1 cm of sil] {}
							edge [pre] node[auto] {} (sil);
							\node [transition] (t2) [right = of p3] {\textbf{A\_FINAL- IZED} 104/104}
							edge [pre] node[auto] {} (p3);
							\node [place] (p4) [ right = of t2] {}
							edge [pre] node[auto] {} (t2);
							
							\node [transition] (t3) [below right = 0.7 and 1 cm of p2]{\textbf{O\_CREA- TED} 104/124}
							edge[pre] node[auto] {} (p2)
							edge[pre] node[auto] {} (p4);
							\node [place] (p5) [right = of t3] {}
							edge[pre] node[auto] {} (t3);
							\node [transition] (t4) [right = of p5] {\textbf{O\_SENT} 104/124}
							edge[pre] node[auto] {} (p5);
							\node [place,pattern=custom north west lines,hatchspread=1.5pt,hatchthickness=0.25pt,hatchcolor=gray] (p6) [right = of t4] {}
							edge[pre] node[auto] {} (t4);
							\end{tikzpicture}}
					}
					\subfloat{
						\raggedleft
						\hspace{-0.22cm}
						\raisebox{1.23\height}{
							\scalebox{0.7}{
								\begin{tabular}{|l|l|}
									\hline
									Support & 0.9905\\
									Confidence& 0.8739\\
									Language fit$_5$& 1.0000\\
									Determinism& 0.8811\\
									Coverage& 0.1723\\
									\hline
								\end{tabular}
							}
						}
					}
				}
				\\
				\vspace{-0.65cm}
				\subfloat{
					\centering
					\subfloat{
						\raggedleft
						\hspace{1.8cm}
						\raisebox{2.9\height}{(d)}
						\scalebox{0.75}{
							\begin{tikzpicture}
							[node distance=1.4cm,
							on grid,>=stealth',
							bend angle=20,
							auto,
							every place/.style= {minimum size=1mm},
						every transition/.style = {minimum size = 3mm, text width=1.5cm,align=center,font=\scriptsize}
							]
							\node [place, tokens = 1] (p){};
							\node [transition] (2) [above right = 0.7 and 1 cm of p] {\textbf{O\_CANC- ELED} 29/34}
							edge [pre] node[auto] {} (p);
							\node [transition] (t1) [below right = 0.7 and 1 cm of p]{\textbf{A\_FINAL- IZED} 95/104}
							edge[pre] node[auto] {} (p);
							\node [place] (p5) [above right = 0.7 and 1 cm of t1] {}
							edge[pre] node[auto] {} (t1)
							edge[pre] node[auto] {} (2);
							\node [transition] (t3) [right = of p5] {\textbf{O\_CREA- TED} 124/124}
							edge[pre] node[auto] {} (p5);
							\node [place] (p6) [right = of t3] {}
							edge[pre] node[auto] {} (t3);
							\node [transition] (t4) [right = of p6] {\textbf{O\_SENT} 124/124}
							edge[pre] node[auto] {} (p6);
							\node [place,pattern=custom north west lines,hatchspread=1.5pt,hatchthickness=0.25pt,hatchcolor=gray] (p7) [right = of t4] {}
							edge[pre] node[auto] {} (t4);
							\end{tikzpicture}}
					}
					\subfloat{
						\hspace{-0.21cm}
						\raggedleft
						\raisebox{1.23\height}{
							\scalebox{0.7}{
								\begin{tabular}{|l|l|}
									\hline
									Support & 0.9920\\
									Confidence& 0.9374\\
									Language fit$_5$& 1.0000\\
									Determinism& 0.7591\\
									Coverage& 0.1397\\
									\hline
								\end{tabular}
							}}
						}
					}
					\\
					\vspace{-0.3cm}
					\subfloat{
						\centering
						\subfloat{
							\raggedleft
							\hspace{3.99cm}
							\raisebox{4.1\height}{(e)}
							\scalebox{0.75}{
								\begin{tikzpicture}
								[node distance=1.4cm,
								on grid,>=stealth',
								bend angle=20,
								auto,
								every place/.style= {minimum size=1mm},
								every transition/.style = {minimum size = 3mm, text width=1.5cm,align=center,font=\scriptsize}
								]
								\node [place, tokens = 1] (p){};
								\node [transition] (2) [above right = of p] {\textbf{A\_ACCE- PTED} 64/104}
								edge [pre] node[auto] {} (p);
								\node [place] (p5) [below right = of 2] {}
								edge[pre] node[auto] {} (2);
								\node [transition] (t3) [below right = of p] {\textbf{W\_Nabel- len offertes} 48/235}
								edge[pre] node[auto] {} (p)
								edge[post] node[auto] {} (p5);
								\node [transition] (t4) [right = of p5] {\textbf{O\_SELEC- TED} 112/124}
								edge[pre] node[auto] {} (p5);
								\node [place,pattern=custom north west lines,hatchspread=1.5pt,hatchthickness=0.25pt,hatchcolor=gray] (p7) [right = of t4] {}
								edge[pre] node[auto] {} (t4);
								\end{tikzpicture}}
						}
						\subfloat{
							\hspace{-0.24cm}
							\raggedleft
							\raisebox{1.23\height}{
								\scalebox{0.7}{
									\begin{tabular}{|l|l|}
										\hline
										Support & 0.9912\\
										Confidence& 0.3933\\
										Language fit$_5$& 1.0000\\
										Determinism& 0.6666\\
										Coverage& 0.2753\\
										\hline
									\end{tabular}
								}}
							}
						}
						\\
						\caption{Five local process models discovered on the BPI'12 log using the technique presented in this paper. Clearly these models provide more insight than Figure \ref{fig:bpic_inductive}.}
						\label{fig:bpic_local_process_models}
						\vspace{-0.1cm}
					\end{figure}
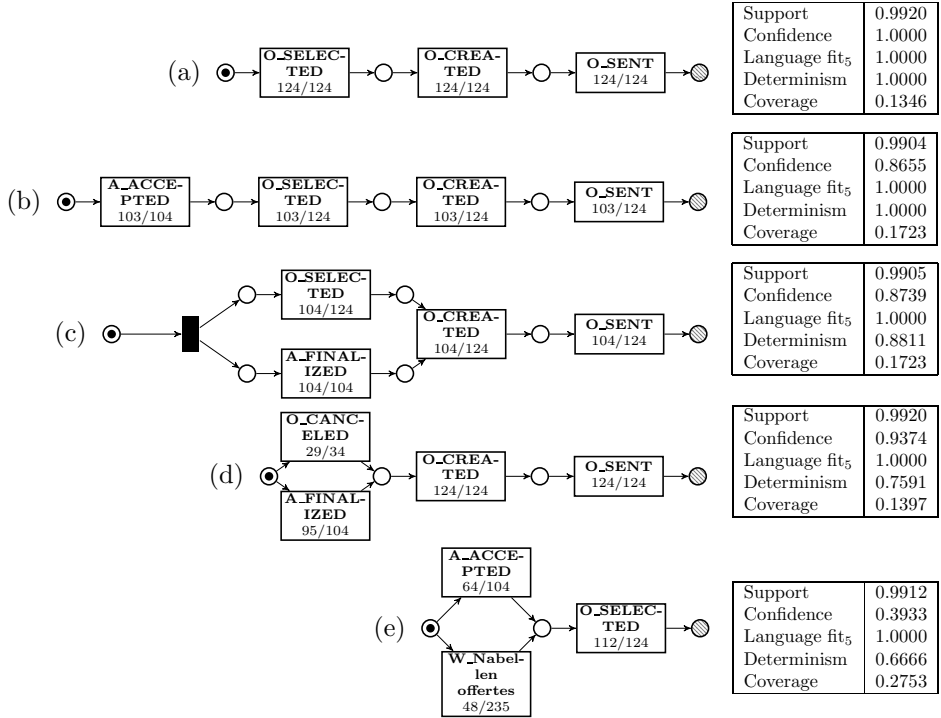

Figure \ref{fig:bpic_inductive} shows the Petri net discovered for resource $10939$ with the Inductive Miner infrequent with a noise threshold of 20\%. The discovered model only contains 13 non-silent transitions, as the activity \emph{W\_valideren aanvraag} is filtered out by the Inductive Miner because of its low frequency. The process model in Figure \ref{fig:bpic_inductive} is very close to a "flower model", which is the model that allows all behavior over its activities. The Inductive Miner without noise filtering returns exactly the flower model over the 14 activities in the log. The discovered process is unstructured because of a high degree of variance of the event log, which is caused by 1) the resource performing work on multiple applications interleaved, and 2) the resource only performing only a subset of the process steps for each application, and which process steps he performs might differ per application. For such a high-variance event log, it is likely that no start-to-end process model exists that accurately describes the behavior in the event log.

Figure \ref{fig:bpic_local_process_models} shows five local process models discovered with the approach described in this paper, which give process insights that cannot be obtained from the start-to-end process model in Figure \ref{fig:bpic_inductive}. Local process model \emph{(a)} shows that all occurrences of events of type \emph{O\_SELECTED}, \emph{O\_CREATED,} and \emph{O\_SENT}, occur in this exact order. Figure \ref{fig:bpic_inductive} overgeneralizes by suggesting that for example \emph{O\_SELECTED} can be followed by three skip (black) transitions, after which another \emph{O\_SELECTED} or a \emph{A\_ACCEPTED} can be performed, which never happens in reality. \emph{O\_SELECTED} and \emph{O\_CREATED} in \ref{fig:bpic_inductive} can be separated by \emph{A\_FINALIZED}, which makes the dependency between \emph{O\_SELECTED} and \emph{O\_CREATED} a long-term dependency, of which discovery is still one of the open problems in process mining \cite{Aalst2004b} The local process model discovery method does find this long term dependency, because each local process model candidate is evaluated on a version of the event log that is projected on the set of labels of candidate under evaluation.

LPM \emph{(b)} is an extension of LPM \emph{(a)} as the last three activities in the sequence are the same, therefore, each occurrence of LPM \emph{(b)} in the log will also be an occurrence of \emph{(a)}. LPM \emph{(b)} starts with an additional activity \emph{A\_ACCEPTED} of which 103 out of 104 events follow this sequential pattern. The confidence of LPM \emph{(b)} is lower than the confidence of \emph{(a)}, because only 103 out of 124 events of the last three activities of the sequence in LPM \emph{(b)} can be explained by the model while each event of these activities is explained by LPM \emph{(a)}. From this we can conclude that there are 21 occurrences of the sequence \emph{O\_SELECTED, O\_CREATED, O\_SENT} that are not preceded by \emph{A\_ACCEPTED}. Partly this can be explained by \emph{A\_ACCEPTED} only occurring 104 times, however, the model also shows that there is one \emph{A\_ACCEPTED} event that is not followed by \emph{O\_SELECTED}, \emph{O\_CREATED}, and \emph{O\_SENT}. It might be the case that this \emph{A\_ACCEPTED} event does not fit the regular workflow, or alternatively it might be the case that the other process steps of after \emph{A\_ACCEPTED} were executed by a different resource. Note that the determinism of LPMs \emph{(a)} and \emph{(b)} is 1.0, since both LPMs are sequential. Language fit of both LPMs is also 1.0, since both allow for only one execution path, which is observed in the log.\looseness=-1

Local process model \emph{(c)} shows that all instances of \emph{A\_FINALIZED} are in parallel with \emph{O\_SELECTED}, and ultimately followed by \emph{O\_CREATED} and \emph{O\_SENT}. This is more informative than Figure \ref{fig:bpic_inductive}, which allows for much more behavior over activities \emph{A\_FINALIZED}, \emph{O\_SELECTED}, \emph{O\_CREATED}, and \emph{O\_SENT}.\looseness=-1

Local process model \emph{(d)} shows that each \emph{O\_CREATED} and \emph{O\_SENT} is preceded by either \emph{O\_CANCELED} (29 times) or \emph{A\_FINALIZED} (95 times). Also most of the \emph{O\_CANCELED} events (29 out of 34) and most of the \emph{A\_FINALIZED} events (95 out of 104) are followed by \emph{O\_CREATED} and \emph{O\_SENT}. Figure \ref{fig:bpic_inductive} does not provide the insight that \emph{O\_CANCELED} is followed by \emph{O\_CREATED} and \emph{O\_SENT}. Note that the determinism of LPM \emph{(d)} is lower than the determinism of LPM \emph{(c)}. This is in agreement with the intuition of determinism, as the concurrency at the start of LPM \emph{(c)} can be regarded as a choice between two activities followed by a deterministic step of executing the other activity, while LPM \emph{(d)} starts with a choice between two activities. After the concurrency in LPM \emph{(c)} and the choice in LPM \emph{(d)} respectively, the two models proceed identically. Local process model \emph{(d)} has higher confidence than LPMs \emph{(b)} and \emph{(c)} as only five of the \emph{O\_CANCELED} and nine of the \emph{A\_FINALIZED} events cannot be explained by the model.
LPM \emph{(d)} has a higher confidence than LPM \emph{(c)}, mostly because all occurrences of \emph{O\_CREATED} and \emph{O\_SENT} could be aligned in LPM \emph{(d)} while only 104 out of 124 could be aligned in LPM \emph{(c)}.\looseness=-1

Notice that the number of events that were aligned on \emph{A\_FINALIZED} is lower in LPM \emph{(d)} than in LPM \emph{(c)}. This indicates that there are six occurrences where the alignments aligned on \emph{O\_CANCELED} while it was possible as well to align on \emph{A\_FINALIZED} (as both occurred). Therefore, an inclusive choice construct would have been a more correct representation than the exclusive choice that is currently included in the LPM. Note that our process tree based discovery approach allows for easy extension with additional operators, like e.g. an inclusive choice operator.\looseness=-1

LPM \emph{(e)} shows an example of a weaker local process model that performs lower on some quality metrics but can still be discovered with the described approach. The coverage of LPM \emph{(e)} is much higher than the other models as \emph{W\_Nabellen offertes} (Dutch for ``Calling after call for bids'') is a frequently occurring event in the log. The confidence of LPM \emph{(e)} is however much lower it explains only a fraction of the \emph{W\_Nabellen offertes} events.

\subsection{Comparison with Related Techniques}
In this section we apply some of the related techniques described in Section \ref{sec:related} to the event log of BPI'12 resource 10939 and compare the insights that can be obtained with those methods with the insights that we obtained with LPM discovery. 

\begin{figure}
	\centering
	\subfloat[]{
		\includegraphics[width=0.65\textwidth]{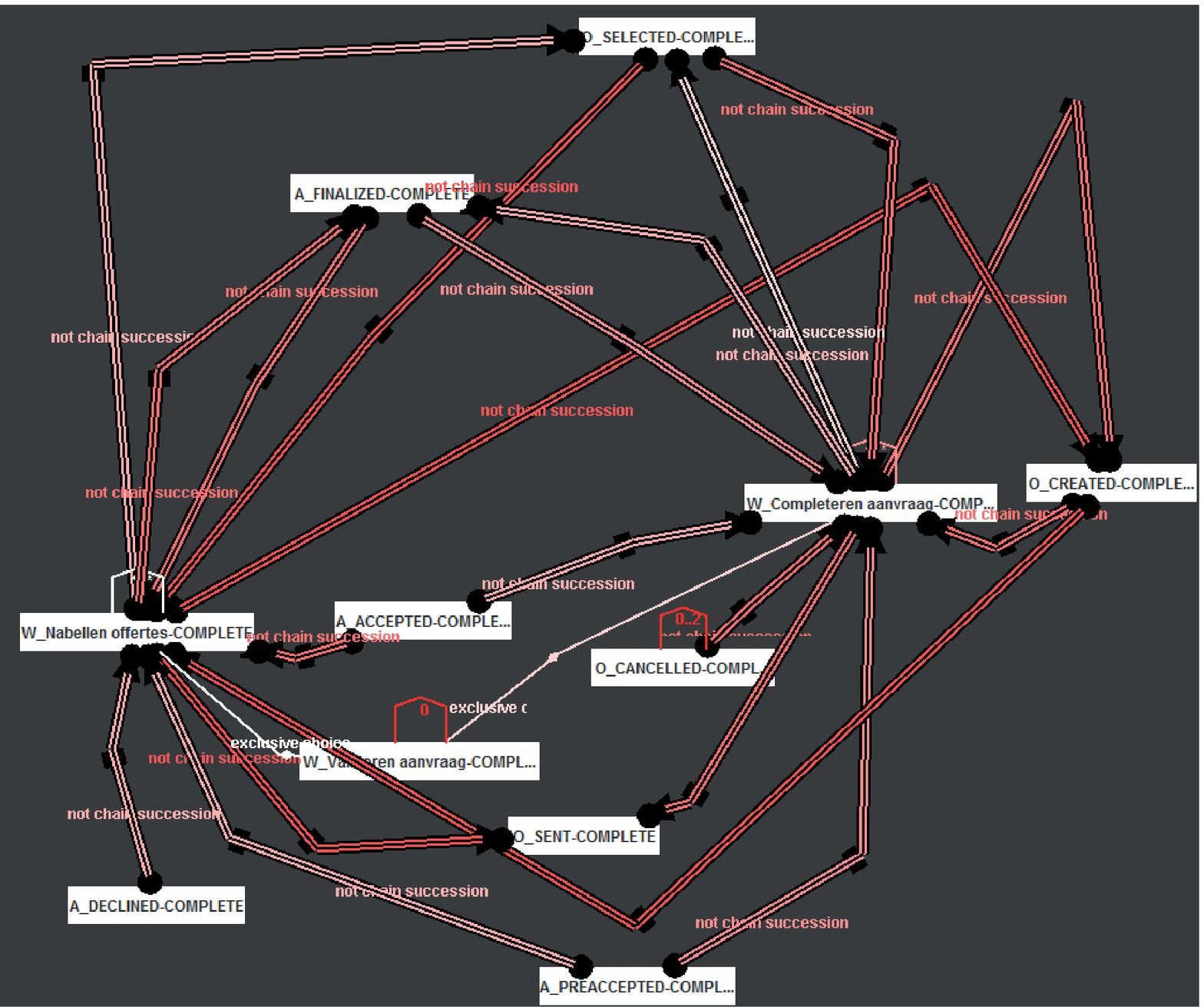}
		\label{sfig:bpi_declare_80}
	}
	\subfloat[]{
		\includegraphics[width=0.35\textwidth]{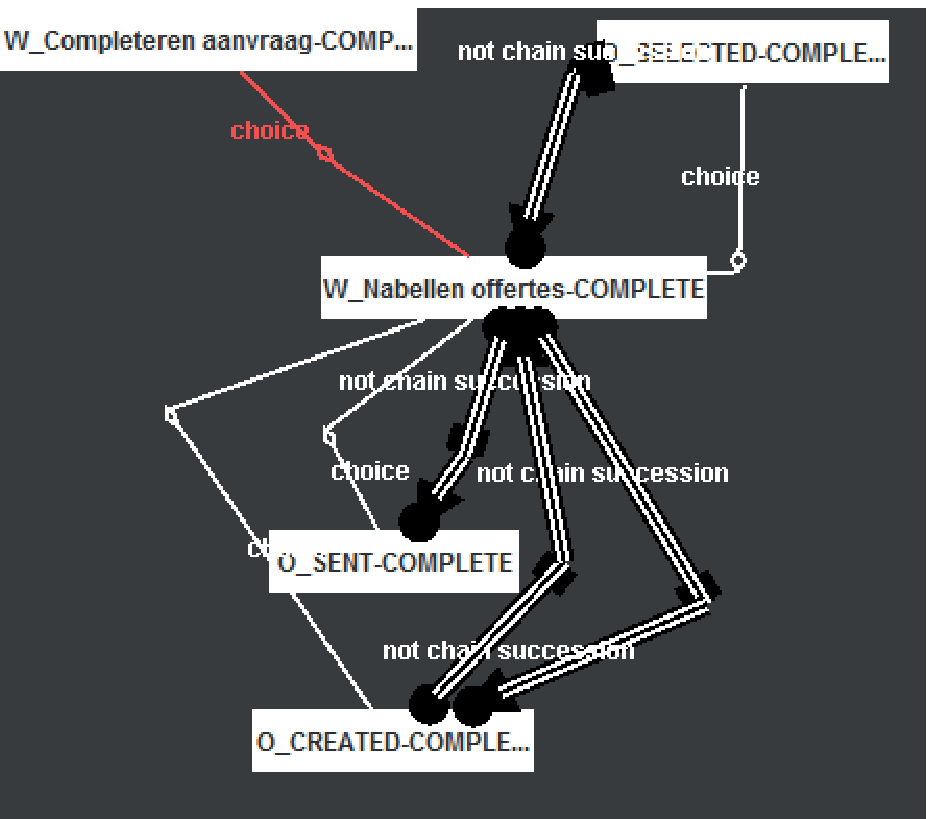}
		\label{sfig:bpi_declare_90}
		
	}
	\caption{\emph{(a)} The result of the Declare miner \cite{Maggi2011} with a support threshold of 80\% on the BPI'12 resource 10939 event log. \emph{(b)} The result of the same method on the same log with support threshold 90\%.}
	\label{fig:bpi_declare}
\end{figure}

We start with the Declare miner \cite{Maggi2011}, which mines a set of binary constraints from the data based on a set of constraint templates. Figure \ref{sfig:bpi_declare_90} shows the result of the Declare miner \cite{Maggi2011} on the BPI'12 resource 10939 event log with a support threshold of 90\%, requiring that the constraints hold in 90\% of the cases. The model shows that a \emph{choice} constraint holds between \emph{O\_SELECTED} and \emph{W\_Nabellen offertes}, indicating that on each working day either at least one event of type \emph{O\_SELECTED} or \emph{W\_Nabellen offertes} occurs. The same can be said about the pairs of event \emph{W\_Nabellen offertes} and \emph{O\_SENT},  \emph{W\_Nabellen offertes} and \emph{O\_CREATED}, and \emph{W\_Nabellen offertes} and \emph{O\_Completeren aanvraag}. Furthermore a \emph{not chain succession} constraint is discovered between \emph{W\_Nabellen offertes} and \emph{O\_SENT}, indicating that \emph{W\_Nabellen offertes} and \emph{O\_SENT} never directly follow each other. \emph{Not chain succession} constraints are also discovered between \emph{W\_Nabellen offertes} and \emph{O\_SELECTED}, and between \emph{W\_Nabellen offertes} and \emph{O\_CREATED}. Note that the none of the insights that we obtained from the LPMs in Figure \ref{fig:bpic_local_process_models} could be obtained from this Declare model.

\begin{figure}
	\centering
	\includegraphics[width=0.5\textwidth]{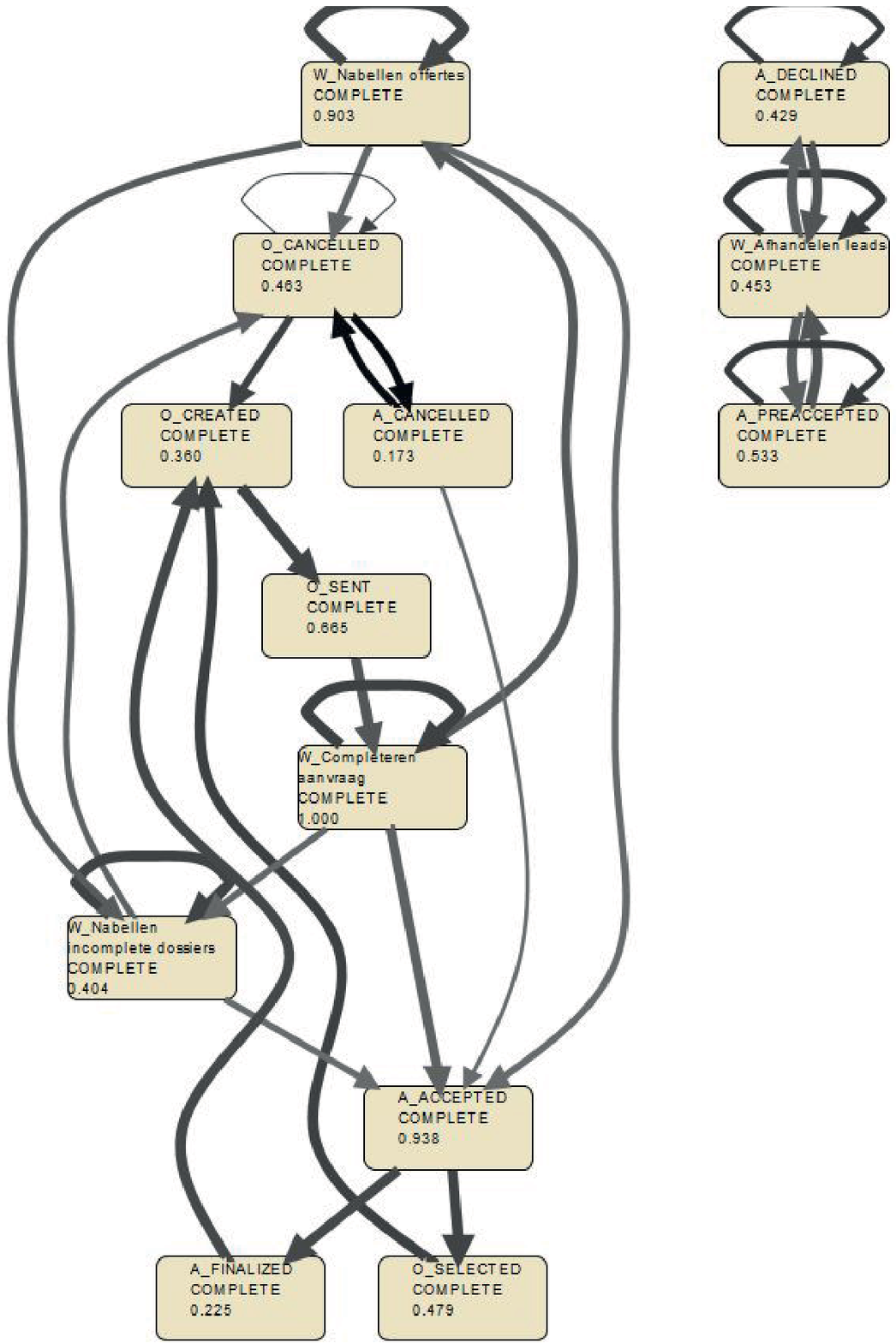}
	\caption{Result of the Fuzzy miner \cite{Gunther2007} with default parameters on the BPI'12 resource 10939 event log.}
	\label{fig:bpi_fuzzy}
\end{figure}

By lowering the support threshold parameter of the Declare miner to 80\%, we obtain a larger set of constraints. An \emph{exclusive choice} constraint is found between \emph{W\_Valideren aanvraag} and \emph{W\_Nabellen offertes}, indicating that 80\% of the cases contain one of the two activities but not both. The same type of constraint is found between \emph{W\_Valideren aanvraag} and \emph{W\_Completeren aanvraag}. The rest of the constraints found are \emph{not chain succession} constraints. 

To find constraints that can be deduced from the LPMs of Figure \ref{fig:bpic_local_process_models}, such as the sequential ordering between \emph{O\_SELECTED} and \emph{O\_CREATED} from LPM \emph{(a)}, the support threshold would need to be lowered even further, leading to an overload of constraints being found by the Declare miner. Declare miner evaluates the constraints based on the ratio of cases in which the constraint holds. However, when activities are often repeated \emph{within} cases, this is not a useful evaluation criterion. Employee 10939 performs most of the activities multiple times during a working day, therefore, to assess whether an activity $a$ is generally followed by an activity $b$ it is more useful to count the ratio of \emph{occurrences} of activity $a$ that are followed by $b$ as in LPM discovery, instead of the number of cases that contain an $a$ event that is followed by a $b$ event. 

Even more important is the fact that Declare miner is limited to binary constraints while LPM discovery mines n-ary relations. That is likely to be the cause of Declare mining not finding any of the LPM relations found in Figure \ref{fig:bpic_local_process_models}. At the same time this difference provides an explanation why Declare mining finds so many uninteresting \emph{not chain succession} constraints between activities: when there are multiple $a$ events in a trace, you are likely to find at least one $a$ that is in a \emph{not chain succession} relation with some activity $b$, leading to a high ratio of traces that fulfill such a \emph{not chain succession} constraint.

\begin{figure}
	\centering
	\subfloat[]{
		\includegraphics[width=0.4\textwidth]{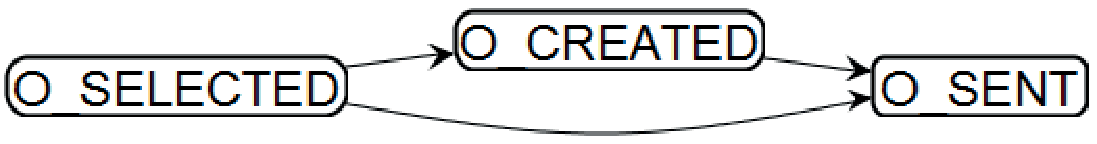}
		\label{sfig:bpi_episode_1}
	}
	\subfloat[]{
		\includegraphics[width=0.4\textwidth]{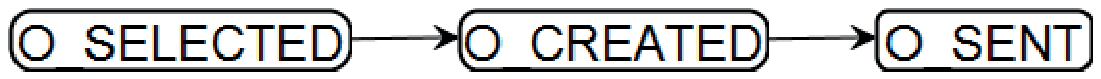}
		\label{sfig:bpi_episode_2}
	}\\
	\subfloat[]{
		\includegraphics[width=0.3\textwidth]{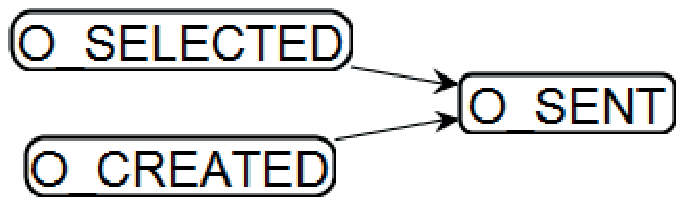}
		\label{sfig:bpi_episode_3}
	}
	\subfloat[]{
		\includegraphics[width=0.3\textwidth]{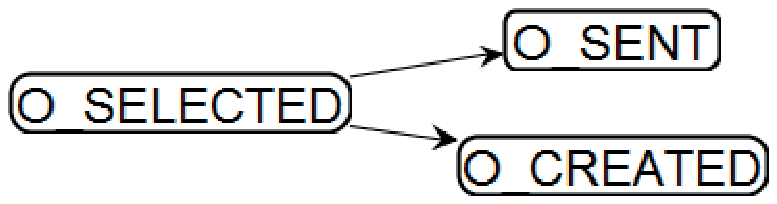}
		\label{sfig:bpi_episode_4}
	}
	\caption{The first four episodes discovered by ProM's Episode Miner \cite{Leemans2014} on the BPI'12 resource 10939 event log.}
	\label{fig:bpi_episodes}
\end{figure}

Figure \ref{fig:bpi_fuzzy} shows the result of the Fuzzy miner on the BPI'12 resource 10939 event log with default parameters. The discovered Fuzzy model does contain a path from \emph{O\_SELECTED} through \emph{O\_CREATED} to \emph{O\_SENT}, which were shown to be in a sequential relation by LPM \emph{(a)}. However, the Fuzzy model allows for many more paths, therefore the sequential relation between those three activities cannot be inferred from the Fuzzy model. LPM \emph{(c)} showed a sequential path between \emph{O\_CREATED} and \emph{O\_SENT} that is preceded by an arbitrary ordering of activities \emph{O\_SELECTED} and \emph{A\_FINALIZED}. The Fuzzy model also shows arrows from both \emph{O\_SELECTED} and \emph{A\_FINALIZED} to \emph{O\_CREATED}, however, as a Fuzzy model does not make a distinction between parallelism, inclusive choice constructs and exclusive choice constructs, it does not answer the question whether \emph{O\_SELECTED} is preceded by \emph{both} \emph{O\_SELECTED} and \emph{A\_FINALIZED}, or just by one of the two.

Figure \ref{fig:bpi_episodes} shows the first four episodes found with ProM's Episode Miner on the BPI'12 resource 10939 event log. The first two episodes show the same sequential ordering from \emph{O\_SELECTED}, \emph{O\_CREATED}, and \emph{O\_SENT} that is represented by LPM \emph{(a)}. The first episode suggests that the \emph{O\_CREATED} event is optional, and can be skipped. LPM \emph{(a)} however shows that all of the \emph{O\_SELECTED} events are followed by an \emph{O\_CREATED} event, therefore it is never skipped. Episode \emph{(c)} indicates that \emph{O\_SELECTED} and \emph{O\_CREATED} can happen in any order, but both of them have to occur before \emph{O\_SENT} occurs and episode \emph{(d)} indicates that \emph{O\_SELECTED} has to happen before \emph{O\_SENT} and \emph{O\_CREATED} can occur. Episodes \emph{(a)}, \emph{(c)} and \emph{(d)} can be considered to be less specific versions of episode \emph{(b)}. ProM's Episode Miner is not able to discover patterns with choice constructs like LPM \emph{(d)}, or patterns with loops.

\subsection{Gazelle Data Set}
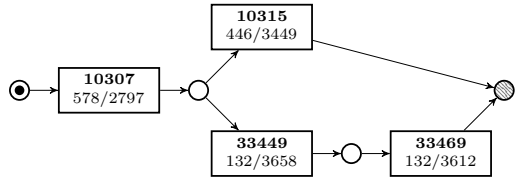
\begin{figure}[th]
	\centering
	\scalebox{0.85}{
		\begin{tikzpicture}
		[node distance=1.4cm,
		on grid,>=stealth',
		bend angle=20,
		auto,
		every place/.style= {minimum size=1mm},
		every transition/.style = {minimum size = 3mm, text width=1.5cm,align=center,font=\scriptsize}
		]
		\node [place, tokens = 1] (p){};
		\node [transition] (t) [right = of p]{\textbf{10307} 578/2797}
		edge [pre] node[auto] {} (p);
		\node [place] (p1) [right = of t]{}
		edge [pre] node[auto] {} (t);
		
		\node [transition] (2) [above right = of p1] {\textbf{10315} 446/3449}
		edge [pre] node[auto] {} (p1);
		\node [transition] (t1) [below right = of p1]{\textbf{33449} 132/3658}
		edge[pre] node[auto] {} (p1);
		\node [place] (p5) [right = of t1] {}
		edge[pre] node[auto] {} (t1);
		\node [transition] (t3) [right = of p5] {\textbf{33469} 132/3612}
		edge[pre] node[auto] {} (p5);
		\node [place,pattern=custom north west lines,hatchspread=1.5pt,hatchthickness=0.25pt,hatchcolor=gray] (p6) [above right = of t3] {}
		edge[pre] node[auto] {} (t3)
		edge[pre] node[auto] {} (2);
		\end{tikzpicture}}
\caption{A non-sequential local process model discovered on the Gazelle data set.\looseness=-1}
\label{fig:gazelle_non_local}
\vspace{-0.23cm}
\end{figure}

The \emph{Gazelle} data set is a real life data set used in the KDD-CUP'2000 and contains customers' web click-stream data provided by the Blue Martini Software company. The Gazelle data set has been frequently used for evaluating sequential pattern mining algorithms. For each customer there is a series of page views, in which each page view is treated as an event. The data set contains 29369 sequences (customers), 87546 events (page views), and 1423 distinct event types (web pages). The average sequence length is three events. More detailed information on the Gazelle data set can be found in \cite{Kohavi2000}. We compare the local process models found on this data set with the sequential patterns obtained with the well-known sequential pattern mining algorithm PrefixSpan \cite{Pei2001} as implemented in the SPMF \cite{Fournier2014} sequential pattern mining library. We set the minimal support parameter of the sequential pattern mining algorithms to 10\% of the number of input sequences. All obtained sequential patterns were also discovered by the local process model miner. Additionally, several non-sequential patterns were discovered that cannot be discovered with sequential pattern mining techniques, an example of which is shown in Figure \ref{fig:gazelle_non_local}. This shows that this well-known sequential pattern mining evaluation data set contains frequent and high-confidence patterns that cannot be found with sequential pattern mining approaches, but can be found with the local process model discovery approach. This indicates the applicability of local process model discovery to the field of pattern mining.



				
		\section{Conclusion \& Future Work}
		\label{sec:conclusion}
		This paper presents a method to discover local process models that can express the same rich set of relations between activities as business process models, but describe frequent fragments instead of complete start-to-end processes. We presented five quality criteria and corresponding metrics quantifying the degree of representativeness of a local process model for an event log. We describe monotonicity properties of quality metrics that can be used to prune the search space and speed up computation. We illustrate through two case studies on real-life data sets that the proposed method enables the user to obtain process insight in the form of valuable patterns when the degree of randomness/variance of the event data prevents traditional process discovery techniques to discover a structured start-to-end process model. Furthermore, the proposed local process model discovery approach is able to discover long-term dependencies, which most process discovery approaches have difficulties with, as a result of evaluating the local process models on a projected version of the event log.
		
		The computational time involved in discovering local process models rapidly grows with the number of activities in the event log. Therefore, we consider automatic discovery of projections on the event log (limiting search to a promising subset of the activities) to be an important area of future work, as it would enable the discovery of local process models on logs with larger numbers of activities. An alternative approach to deal with larger numbers of activities that is to be explored is the use of meta-heuristic search methods, e.g. simulated annealing, which allows partial exploration of the search space.
		
		Finally, we consider it to be a relevant future direction of research to enhance local process models with guards, time information, and resource information.

\bibliography{bare_conf}

\end{document}